\documentstyle[floats,tighten,epsf,preprint,aps]{revtex}
\preprint{KUNS 1213}
\date{August 2, 1993}
\draft
\newcommand\carbon{{}^{12}{\rm C}}
\begin{document} \title{
  Nucleon Flow and Fragment Flow in Heavy Ion Reactions
}
\author{Akira Ono, Hisashi Horiuchi
and Toshiki Maruyama\thanks{Present address:
  Advanced Science Research Center,
  Japan Atomic Energy Research Institute,
  Tokai-mura, Ibaraki-ken 319-11, Japan
}}
\address{
  Department of Physics, Kyoto University, Kyoto 606-01, Japan
}
\maketitle
\begin{abstract}
The collective flow of nucleons and that of fragments in the
$\carbon+\carbon$ reaction below 150 MeV/nucleon are calculated with
the antisymmetrized version of molecular dynamics combined with the
statistical decay calculation.  Density dependent Gogny force is used
as the effective interaction.  The calculated balance energy is about
100 MeV/nucleon, which is close to the observed value.  Below the
balance energy, the absolute value of the fragment flow is larger than
that of nucleon flow, which is also in accordance with data.  The
dependence of the flow on the stochastic collision cross section and
its origin are discussed.  All the results are naturally understood by
introducing the concept of two components of flow: the flow of
dynamically emitted nucleons and the flow of the nuclear matter which
contributes to both the flow of fragments and the flow of nucleons due
to the statistical decay.
\end{abstract}
\narrowtext

\section{Introduction}
The collective flow in heavy ion reactions in the intermediate and
high energy region has been studied in these years both experimentally
and theoretically with the expectation that it carries the information
of hot and dense nuclear matter.  However, by the study with
microscopic simulation approaches such as Vlasov-Uehling-Uhlenbeck
(VUU) method, it has turned out that extracting the equation of state
from the collective flow is not easy since it reflects not only the
equation of state but also the cross section of the two-nucleon
collision term which has theoretical ambiguity in the nuclear medium.

Although the collective flow is a one-body observable, the most
characteristic feature of intermediate energy heavy ion reactions is
the fragment formation.  Recently the collective flow has become to be
measured with the identification of charges and/or mass numbers of
fragments \cite{PETER,WESTFALL}.  The large flow of fragments is
observed compared to the flow of nucleons, which suggests that the
flow of composite fragments carries the direct information of the
equation of state (or the mean field), while most nucleons are emitted
by the hard stochastic collisions which erase the effect of the mean
field.  Thus we can expect that the study of the fragment flow
together with the nucleon flow may give us important information on
the equation of state.

For the theoretical analysis of the flow of fragments, the model
should be able to describe the dynamical fragment formation, and we
use the antisymmetrized version of molecular dynamics (AMD)
\cite{ONOa,ONOb,ONOc} which was constructed by incorporating the
stochastic collision process into the fermionic molecular dynamics
that describes the system with a Slater determinant of Gaussian wave
packets \cite{FELD,HORI,HORIa}.  We have demonstrated that the AMD can
describe some quantum mechanical features such as the shell effect in
the cross sections of the dynamical production of the fragments.
Since the flow of $\alpha$ particles is important among the flows of
various fragments, this ability of the AMD to describe the large cross
sections of the dynamical production of $\alpha$ particles is a very
advantageous property of the AMD for the present study.  Furthermore,
AMD has an advantage that the calculation with finite-range effective
interaction is as feasible as the calculation with zero-range
effective interaction, and therefore the momentum dependence of the
mean field, which is essential for the study of the equation of state,
is automatically taken into the calculation.

In studying the collective flow, we should also pay attention to the
momentum distribution of the produced fragments.  Namely, since the
flow is known to be a very sensitive quantity to the various
parameters in the theory, such dependence should be understood on the
level of the momentum distribution.  For example, some change of the
equation of state and another change of the in-medium cross section
may result in the same change of the flow value, but this does not
mean that one must give up the determination of the equation of state
by using heavy ion reactions.  We still have chance that these two
changes will result in the distinguishable changes of the momentum
distribution of fragments.  Our framework AMD is suitable for such
study of the momentum distribution of fragments since there is no
ambiguity in identifying the fragments.  As we pointed out in Ref.\
\cite{ONOc}, the dissipated component of the momentum distribution of
fragments such as $\alpha$ particles is sensitive to the in-medium
cross section, while the flow angle is expected to be sensitive to the
effective interaction.

In this paper, we will investigate the flow of nucleons and fragments
for the reaction ${}^{12}{\rm C}+{}^{12}{\rm C}$ in the energy region
45 MeV/nucleon $\le E\le$ 150 MeV/nucleon which includes the observed
balance energy.  After giving a brief explanation of our framework in
Sect.\ II, we discuss in Sect.\ III the use of the finite range Gogny
force \cite{GOGNY} as the effective interaction which gives the soft
equation of state with the incompressibility of the nuclear matter
$K=228$ MeV.  It is shown that the Gogny force reproduces binding
energies of light nuclei relevant to the present study.  In Sect. IV,
before showing the calculated results of the flow, we will check the
validity of our calculation by comparing the proton spectra with the
data and the results of other calculations.  In Sect.\ V, the
calculated results of the flow are shown and it will turn out that the
soft equation of state with the Gogny force is consistent to the
experimentally observed balance energy of the flow.  The calculated
flow of $\alpha$ particles is larger than the flow of nucleons, which
is also consistent to the feature of the experimental data, and the
mechanism of the creation of these flows will be revealed by paying
attention to the time scale of the reaction and dividing the
collective flow into two components, i.e., the flow of dynamically
emitted nucleons and the flow of the nuclear matter.  In Sect.\ VI,
the dependence of the flow on the in-medium cross section and the
origin of this dependence are discussed.  Sect.\ VII is devoted to
the summary.

\section{Framework}
Since the framework of the antisymmetrized version of molecular
dynamics (AMD) was described in detail in Refs.\ \cite{ONOb,ONOc},
here is shown only the outline of our framework.

In AMD, the wave function of $A$-nucleon system $|\Phi\rangle$ is 
described by a Slater determinant $|\Phi({\bf Z})\rangle$;
\begin{equation}
  |\Phi({\bf Z})\rangle=
   {1\over\sqrt{A!}}\det\Bigl[\varphi_i(j)\Bigr],\quad
   \varphi_i=\phi_{{\bf Z}_i}\chi_{\alpha_i},
\end{equation}
where $\alpha_i$ represents the spin-isospin label of $i$th single
particle state, $\alpha_i={\rm p}\uparrow$, ${\rm p}\downarrow$, ${\rm
n}\uparrow$, or ${\rm n}\downarrow$, and $\chi$ is the spin-isospin
wave function. $\phi_{{\bf Z}_i}$ is the spatial wave function of
$i$th single particle state which is a Gaussian wave packet
\begin{equation}
  \langle{\bf r}\,|\phi_{{\bf Z}_i}\rangle=
  \Bigl({2\nu\over\pi}\Bigr)^{3/4}
  \exp\biggl[-\nu\Bigl({\bf r}-{{\bf Z}_{i}\over\sqrt\nu}\Bigr)^2
                      +{\textstyle{1\over2}}{{\bf Z}}_{i}^2\biggr],
\end{equation}
where the width parameter $\nu$ is treated as time-independent in our
model.  We took $\nu=0.16$ ${\rm fm}^{-2}$ in the calculation
presented in this paper.

The time developments of the centers of Gaussian wave packets, $\{{\bf
Z}_i (i=1,2,\ldots,A)\}$, are determined by two processes.  One is the
time development determined by the time-dependent variational
principle
\begin{equation}
  \delta\int_{t_1}^{t_2}dt\,
  { \langle\Phi({\bf Z})|(i\hbar{d\over dt}-H)|\Phi({\bf Z})\rangle
   \over\langle\Phi({\bf Z})|\Phi({\bf Z})\rangle}=0,
\end{equation}
which leads to the equation of motion for $\{{\bf Z}\}$.

The second process which determines the time development of the system
is the stochastic collision process due to the residual interaction.
We incorporate this process in the similar way to the quantum
molecular dynamics (QMD) by introducing the physical coordinates
$\{{\bf W}_i\}$ \cite{ONOa,ONOb} as
\begin{equation}
  {\bf W}_i=\sum_{j=1}^A \Bigl(\sqrt Q\Bigr)_{ij}{\bf Z}_j,
\end{equation}
where
\begin{equation}
  Q_{ij}={\partial\over\partial({\bf Z}_i^*\cdot{\bf Z}_j)}
         \log \langle\Phi({\bf Z})|\Phi({\bf Z})\rangle.
\end{equation}
In addition to the usual two nucleon collision process, the stochastic
nucleon-alpha collision process is also included as in Ref.\
\cite{ONOc}.  Since we apply AMD to the reactions with higher incident
energy in this paper than in the previous works \cite{ONOa,ONOb,ONOc},
the cross sections and the angular distributions of the stochastic
collisions are reparametrized as explained in the Appendix B.

Since in AMD the center-of-mass motion of the fragment is described by
the Gaussian wave packet, we have to subtract from the AMD Hamiltonian
$\langle\Phi({\bf Z})|H|\Phi({\bf Z})\rangle/\langle\Phi({\bf
Z})|\Phi({\bf Z})\rangle$ the sum of the spurious zero-point energies
of the fragments whose number changes in time.  The prescription to
deal with this problem which is a revised version of the previous
prescription in Refs.\ \cite{ONOa,ONOb,ONOc} is explained in the
Appendix C.

The simulations of AMD are truncated at a finite time (150 fm/$c$ in
the calculation presented in this paper).  The dynamical stage of the
reaction has finished by this time and some excited fragments have
been formed which will emit lighter particles with a long time scale.
Such statistical decays of the equilibrated fragments are calculated
with a code of Ref.\ \cite{MARUb} which is similar to the code of
P\"uhlhofer \cite{PUHLHOFER}.

\section{Choice of the effective interaction}

Since the collective flow is expected to reflect the momentum
dependence and the density dependence of the mean field, the choice of
the effective interaction in our framework is important.

\begin{figure}
\centerline{\epsfysize=9cm \epsffile{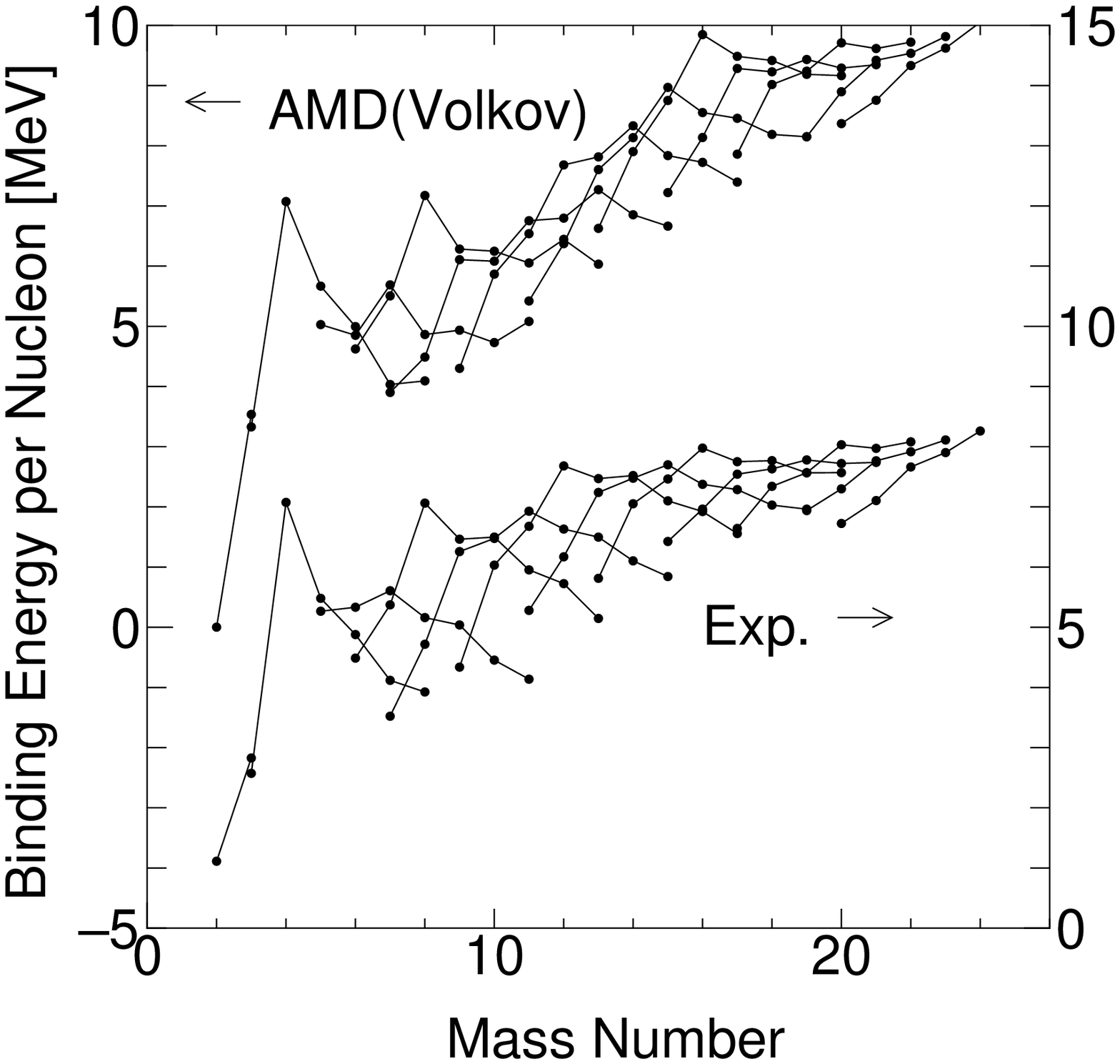}
            \epsfysize=9cm \epsffile{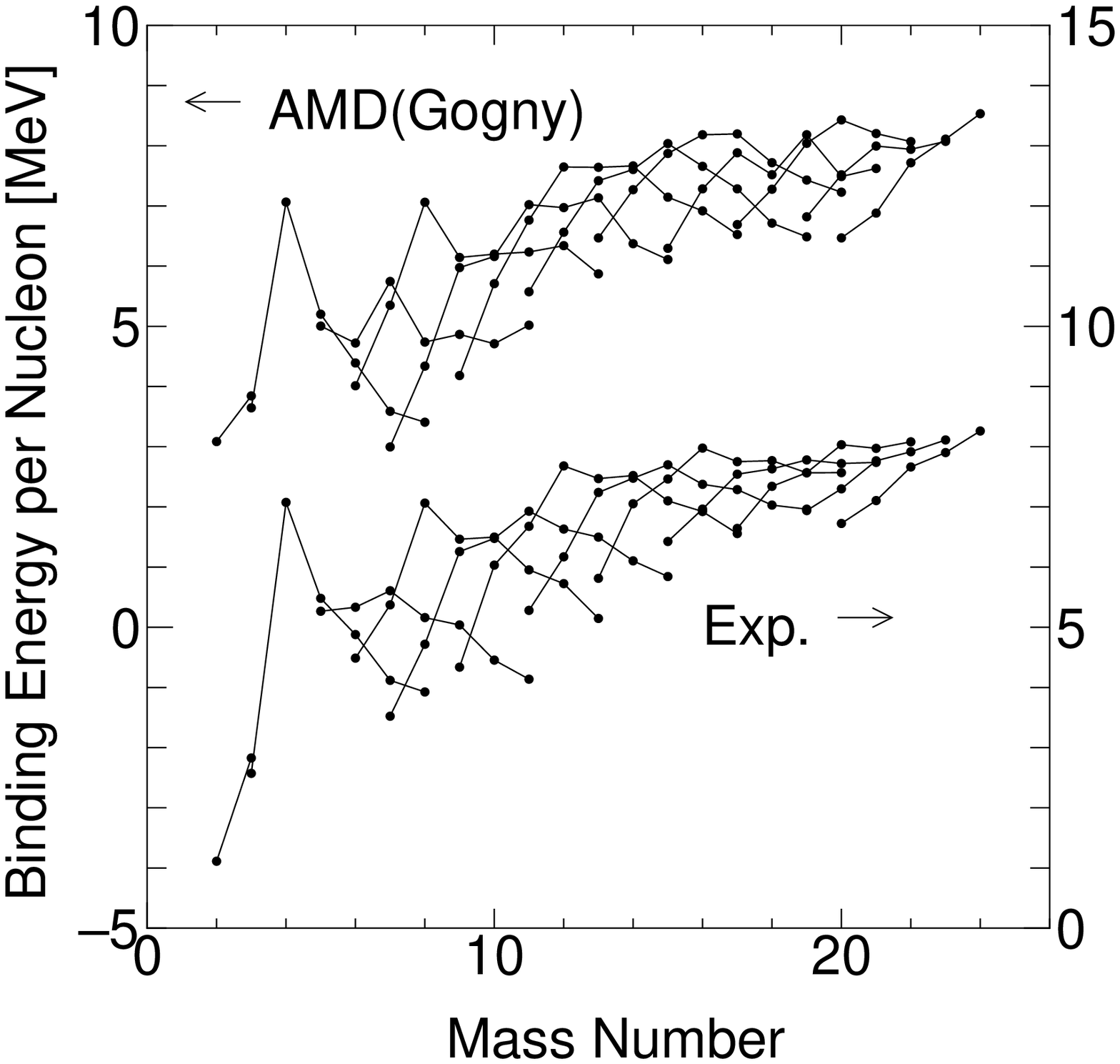}}
\vspace{-0.5cm}
\caption{\label{FigBindingEnergies}
Binding energies per nucleon of nuclei calculated with the Volkov
force No.\ 1 ($m=0.576$) (left) and the Gogny force (right).  Isotopes
are connected with lines.  Experimental data shifted by $-5$ MeV are
also shown.  }
\end{figure}

The momentum dependence of the mean field can be taken into account as
long as a finite range effective interaction is used, since the system
is described with an antisymmetrized wave function in AMD.  The
calculation with a finite range force is as feasible as the
calculation with a zero-range force in AMD, and therefore we have been
using finite range effective interactions.  In Refs.\
\cite{ONOa,ONOb,ONOc} we used the Volkov force No.\ 1 \cite{VOLKOV}
with the Majorana parameter $m=0.576$, which is a density-independent
two-range interaction.  By adjusting the parameter related to the
subtraction of the spurious zero-point oscillation of fragments, the
binding energies of nuclei lighter than $^{12}{\rm C}$ are reproduced
very well.  However, heavier nuclei are overbinding with this
effective interaction as shown in Fig.\ \ref{FigBindingEnergies}.
When the Volkov force is applied, usually its Majorana parameter $m$
should be chosen depending on the mass number of the nuclei under
consideration, which makes the Volkov force very inconvenient for our
study.

In order to reproduce the binding energies of nuclei in wide mass
number region within a single effective interaction, and in order to
cope duely with the formation of the high density nuclear matter in
collisions with high incident energy around and above 100 MeV/nucleon,
the density dependence of the effective interaction should be
introduced.  We use the Gogny force \cite{GOGNY} in the calculation
presented in this paper.  The binding energies of nuclei lighter than
${}^{24}{\rm Mg}$ are reproduced (Fig.\ \ref{FigBindingEnergies}) by
adjusting the parameter related to the subtraction of the spurious
zero-point oscillation of fragments which is described in the Appendix
C.  It is known that the Gogny force can describe the saturation
property of nuclear matter with the incompressibility $K=228$ MeV.
The momentum dependence of the mean field with the Gogny force is
reliable below 200 MeV, since the energy dependence of the real part
of the nucleon optical potential is reproduced in this energy range.
The Gogny force has ever been successfully applied to the problem of
the flow in the framework of VUU \cite{SEBILLE}.

The calculation with density dependent zero-range force requires us to
evaluate the quantities such as
\begin{equation}
\int d{\bf r}\,\rho({\bf r})^{2+\sigma},
\end{equation}
where $\rho({\bf r})$ is the density calculated with the AMD wave
function.  This spatial integration was performed by a kind of Monte
Carlo method with test particles generated randomly with a weight
function.  The detail of the calculation method is explained in the
Appendix A.  If we use $100A$ test particles, the statistical error in
the evaluation of the energy of $\carbon$ ground state is about 2 MeV,
which seems sufficiently small for the current purpose.

\section{Energy spectra of protons}
Before going into the problem of the collective flow, we compare our
results on the proton energy spectrum to the data in the reaction
$^{12}{\rm C}+{}^{12}{\rm C}$ reaction at the incident energy 84
MeV/nucleon, which was also analyzed by other simulation methods such
as VUU and QMD \cite{AICH}.  It is for the purpose of showing the
reliability of the AMD calculation in the incident energy region
around 100 MeV/nucleon.  As mentioned before \cite{ONOa,ONOb,ONOc},
the AMD has proved to be very successful in reproducing the nuclide
distribution and the fragment energy spectra in the incident energy
region around the Fermi energy.

\begin{figure}
\epsfysize=9cm
\centerline{\epsffile{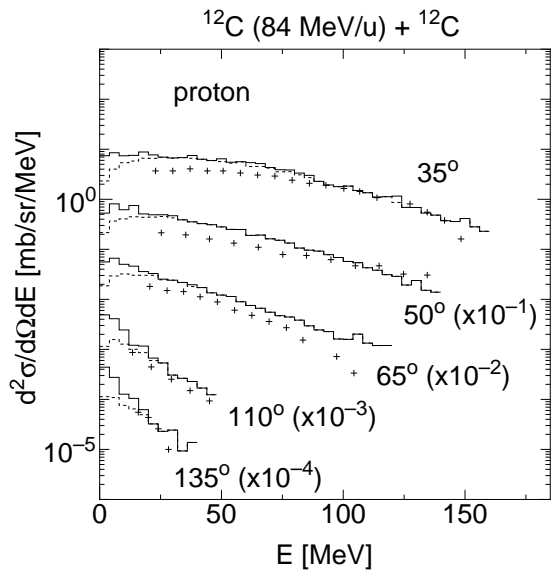}}
\vspace{-0.5cm}
\caption{\label{FigProtonSpectrum}
Energy spectra of protons for the reaction $\carbon+\carbon$ at 84
MeV/nucleon.  Dashed histograms are the proton spectra calculated
before the statistical decay, and the solid histograms are the final
calculated results.  Experimental data are shown by crosses.  Energy
$E$ and angles are in the laboratory system.  }
\end{figure}

\begin{figure}
\epsfysize=9cm
\centerline{\epsffile{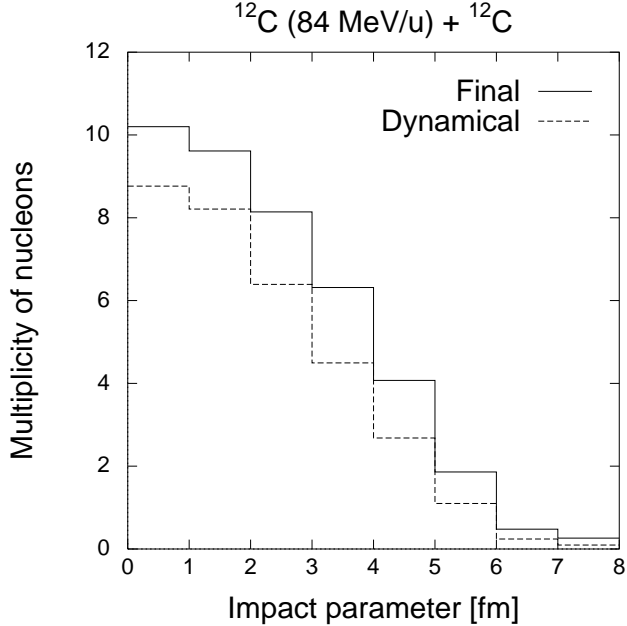}}
\caption{\label{FigMultNuc}
Impact parameter dependence of the nucleon multiplicity for the
reaction $\carbon+\carbon$ at 84 MeV/nucleon.  Dashed histogram shows
the multiplicity before the statistical decay and the solid histogram
shows the final calculated result.  }
\end{figure}

In Fig.\ \ref{FigProtonSpectrum}, the calculated proton energy spectra
for various angles are compared to the data \cite{BRUMMUND}.  The
dotted histograms are the calculated proton spectra at the end of AMD
simulation $t=150$ fm/$c$, which do not contain the contribution from
the statistical decay of excited fragments after $t=150$ fm/$c$, while
the solid histograms are the final results which should be compared to
the data.  We can see the effect of the statistical decay of fragments
on these spectra is not so large except for the low energy part.
Although our calculation has tendency to overestimate the proton
spectra at $35^\circ$--$65^\circ$, this feature and the degree of the
overestimation are common with other simulation approaches such as VUU
and QMD \cite{AICH}.  The spectra of backward angles are reproduced by
our model as well as by QMD.  In Fig.\ \ref{FigMultNuc}, the
multiplicity of nucleons are shown as a function of the impact
parameter.  The dynamical contribution before the statistical decay is
shown as well as the final multiplicity.  This result is again quite
similar to those of QMD and VUU \cite{AICH} for all the impact
parameter region.

In calculating the spectra in our model, we must attribute some widths
to the momenta of nucleons and fragments produced in the AMD
calculation, since these momenta are central values of Gaussian wave
packets.  Each wave packet has the width $\Delta p/\sqrt{A_{\rm F}}$
in the momentum per nucleon where $\Delta p=\hbar\sqrt{\nu}$ and
$A_{\rm F}$ is the mass number of the fragment.  However, we should
not always take this width to have physical meaning, since some part
of this width comes from the unphysical width of the projectile or the
target in the initial state of the simulation.  It seems reasonable to
consider that such unphysical contribution mainly goes to the momentum
widths of fragments and to assume that all the widths of the wave
packets of nucleons are physical.  When the proton spectra in Fig.\
\ref{FigProtonSpectrum} are calculated, we neglect the
widths of the fragments produced dynamically in the AMD simulation,
but attribute the widths to the dynamically produced nucleons in the
AMD simulation.

\section{Calculated results of the flow
and the mechanism of the flow creation}
Although the definition of the flow is somewhat complicated in the
experimental situation where the determination of the reaction plane
is not trivial, we simply define here the flow of fragments with mass
number $A_{\rm F}$ as the mean in-plane transverse momentum,
\begin{equation}
\langle wP_x\rangle/A_{\rm F},
\label{EqDefFlow}
\end{equation}
where $w=1$ if $P_z>0$ and $w=-1$ if $P_z<0$.  $P_x$ and $P_z$ are the
components of the in-plane momentum of the fragment in the
center-of-mass system which are perpendicular to and along with the
beam direction respectively.  Negative flow corresponds to the
attractive interaction between the projectile and the target.  This
definition of the flow was already used in Ref.\ \cite{KLAKOW} and has
almost no uncertainty due to the interpretation of the momentum width
of the wave packets discussed in the previous section.

Another definition of the flow as the slope of $\langle
P_x\rangle$-$P_z$ curve is of course possible, but we have checked
that the qualitative features to be discussed below does not change.
The statistical precision of the flow defined by Eq.\
(\ref{EqDefFlow}) is better than the latter one.

\begin{figure}
\epsfysize=9cm
\centerline{\epsffile{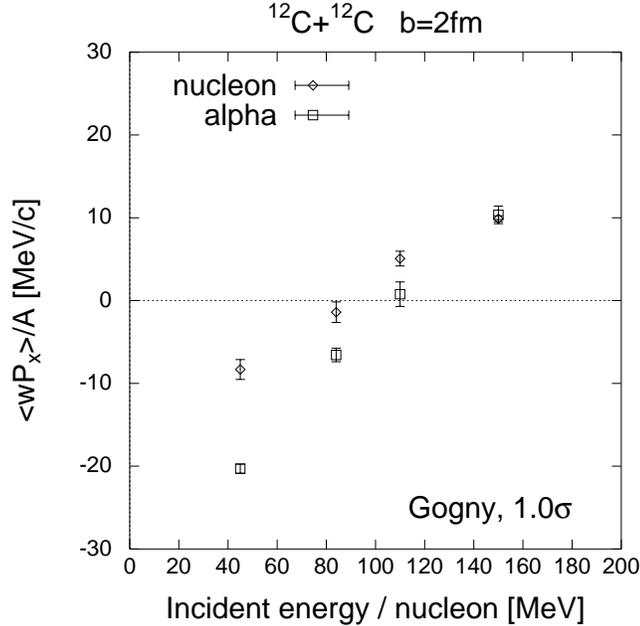}}
\caption{\label{FigFlowEng}
Calculated flows of nucleons and $\alpha$ particles for various
incident energies for the reaction $\carbon+\carbon$.  The impact
parameter of the simulation is fixed to be 2 fm.  }
\end{figure}

Fig.\ \ref{FigFlowEng} shows the energy dependence of the calculated
flows of nucleons and $\alpha$ particles.  In this figure, as well as
in all the results hereafter, the impact parameter of the reaction is
fixed to be 2 fm in order to improve the statistical precision.  We
can see that the calculated flows of nucleons and $\alpha$ particles
change their sign at the balance energy $E_{\rm bal}= (100\pm20)$ MeV,
though the balance energy of $\alpha$ particles seems to be larger
than that of nucleons.  This calculated balance energy is not
different so much from the experimentally deduced value $122\pm12$
MeV/nucleon \cite{WESTFALL}.  On the other hand, in the calculation
with the Volkov force, the flows of nucleons and $\alpha$ particles
are negative even at the incident energy 150 MeV/nucleon
\cite{ONOrcnpanu}. (In this calculation, another version of the
stochastic collision process adopted in Ref.\ \cite{ONOc} was used.)

\begin{figure}
\epsfysize=9cm
\centerline{\epsffile{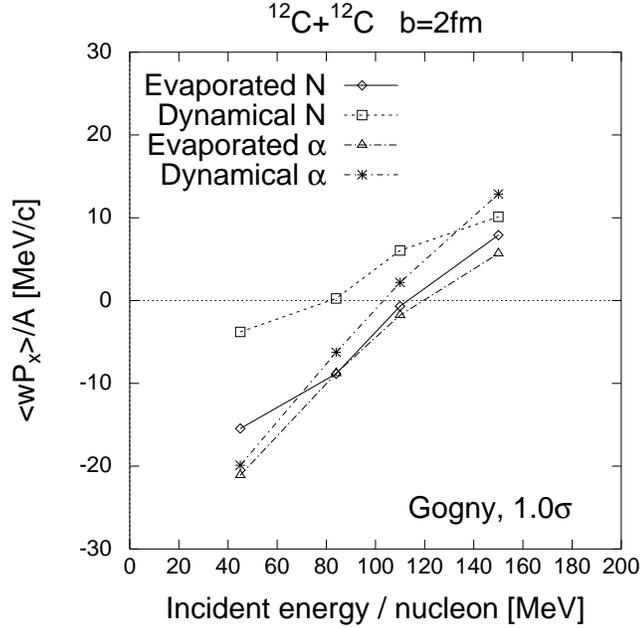}}
\caption{\label{FigFlowEngDynEvap}
The dynamical flows of nucleons and $\alpha$ particles which are
calculated before the statistical decay, and the flows of evaporated
particles which are calculated only from the decay products of the
statistical decay.  }
\end{figure}

\begin{figure}
\epsfysize=9cm
\centerline{\epsffile{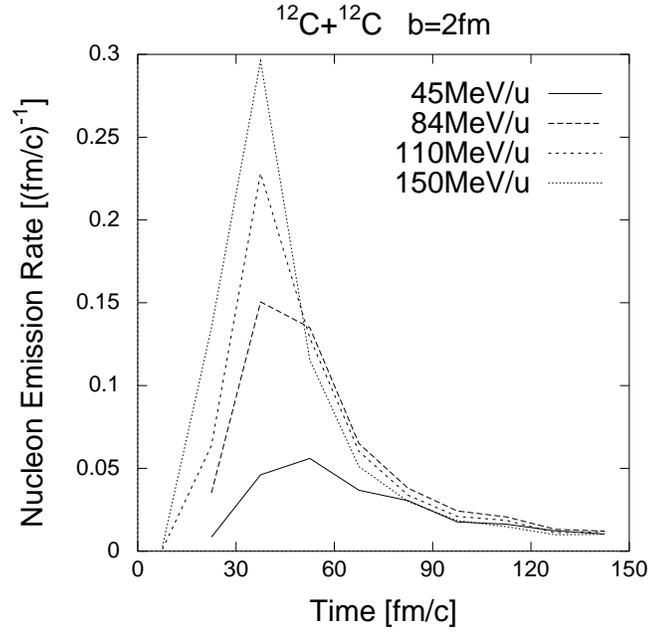}}
\caption{\label{FigdNdt}
Time dependence of the nucleon emission rate for the central reaction
$\carbon+\carbon$ ($b=2$ fm) at various incident energies.  A nucleon
is interpreted to have been emitted when no other nucleons exist
within a radius of 3 fm.  }
\end{figure}

\begin{figure}
\epsfysize=9cm
\centerline{\epsffile{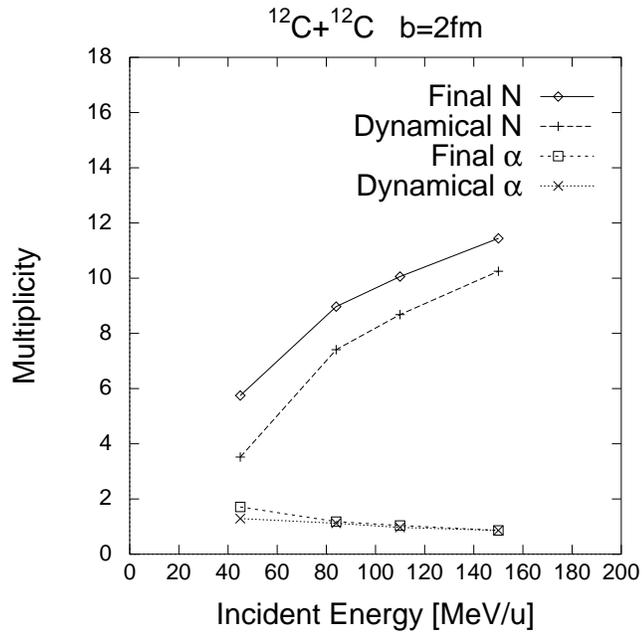}}
\caption{\label{FigMultEng}
Incident energy dependence of the multiplicities of nucleons and
$\alpha$ particles for the reaction $\carbon+\carbon$ with $b=2$ fm.
It is to be noted that some of dynamical $\alpha$ particles decay in
the long time scale.
}
\end{figure}

A remarkable feature is that the absolute value of the flow of
$\alpha$ particles is much larger than that of nucleons in the energy
region $E<E_{\rm bal}$.  The reason of this larger flow of fragments
becomes clear by distinguishing nucleons and fragments according to
the time they are produced.  From Fig.\ \ref{FigdNdt} which shows the
distribution of the emission time of nucleons, we can see that the
dynamical stage of the reaction has already finished at $t=150$
fm/$c$.  Fig.\ \ref{FigFlowEngDynEvap} shows the flows at $t=150$
fm/$c$ (Dynamical) and the flows of nucleons and $\alpha$ particles
which are produced by the statistical decay of other fragments after
$t=150$ fm/$c$ (Evaporation).  Although the `dynamical' flows in the
figure contain some contribution from the equilibrated stage of the
reaction, we can see clear deference between the flow of dynamically
emitted nucleons and the flow of evaporated nucleons from the
equilibrated fragments.  The negative flow of the evaporated nucleons
is much larger than the flow of the dynamically emitted nucleons, and
the final result is between them.  Since larger number of nucleons are
emitted in the dynamical stage, as can be seen from Fig.\
\ref{FigMultEng}, the final result is closer to the flow of the
dynamically emitted nucleons.  On the other hand, the flow of
$\alpha$ particles seems to be almost independent of the time they are
produced, and the value of the flow is close to the flow of the
evaporated nucleons.

These results lead us almost uniquely to the following interpretation.
We can separate the collective flow into two components.  The first
one is the flow of the nucleons which are emitted by the hard
stochastic collisions in the dynamical stage of the reaction.  Since
the stochastic collision erases the memory of the mean field which the
nucleon has been feeling before it is scattered, the absolute value of
the negative flow is small.  There may be another geometrical reason
for this small flow that the projectile and the target prevent the
nucleons with negative flow from going out to the free space more than
the nucleons with positive flow.  The second component of the
collective flow is the flow of the nuclear matter which contributes to
the flows of composite fragments produced in the dynamical stage of
the reaction.  This component has been affected directly by the mean
field and the absolute value of the negative flow is large.  The flows
of evaporated nucleons and fragments inherit the flow of the parent
fragments and hence they have the large value similar to the original
nuclear matter flow.

\section{Dependence on the stochastic collision cross section}
Since the stochastic collision cross sections have some theoretical
uncertainty, we should investigate the dependence of the flow on them.
Furthermore the study of such $\sigma$-dependence of the momentum
distribution will give us the way to determine the in-medium cross
sections.

\begin{figure}
\centerline{\epsfysize=9cm \epsffile{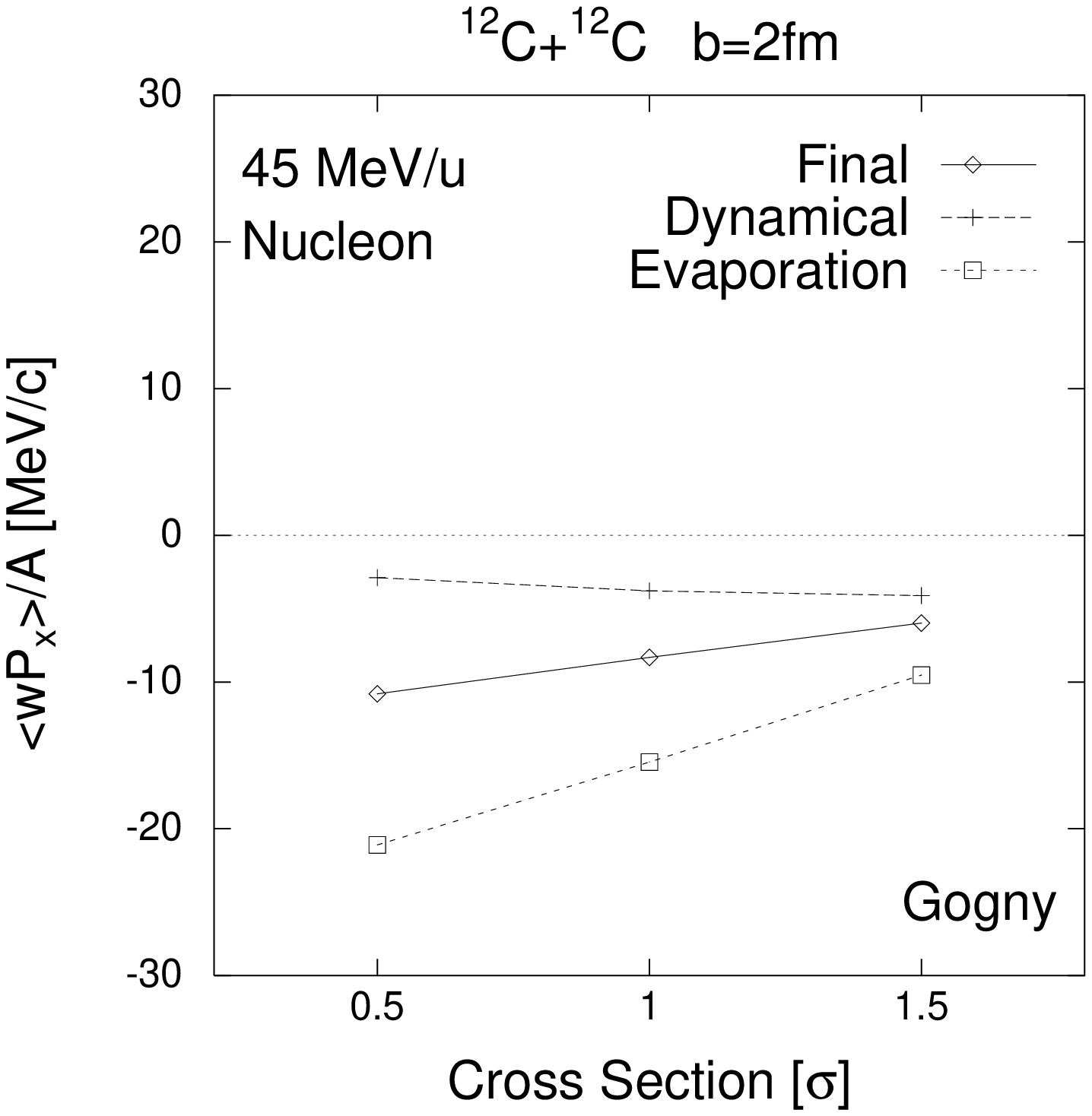}
            \epsfysize=9cm \epsffile{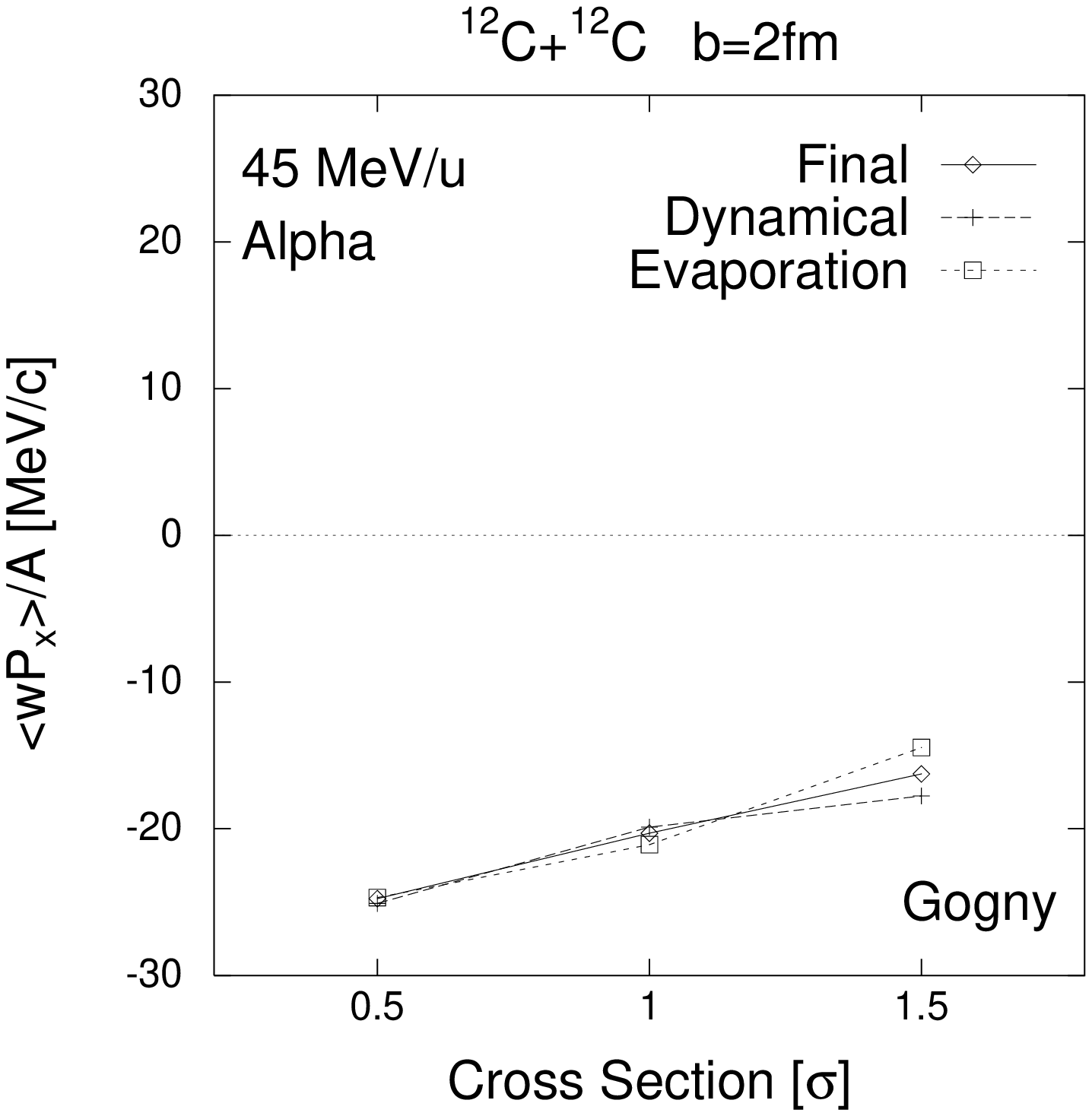}}
\caption{\label{FigFlowSig}
Dependence of the flows on the stochastic collision cross sections for
the reaction $\carbon+\carbon$ at 45 MeV/nucleon.  }
\end{figure}

\begin{figure}
\centerline{\epsfysize=9cm \epsffile{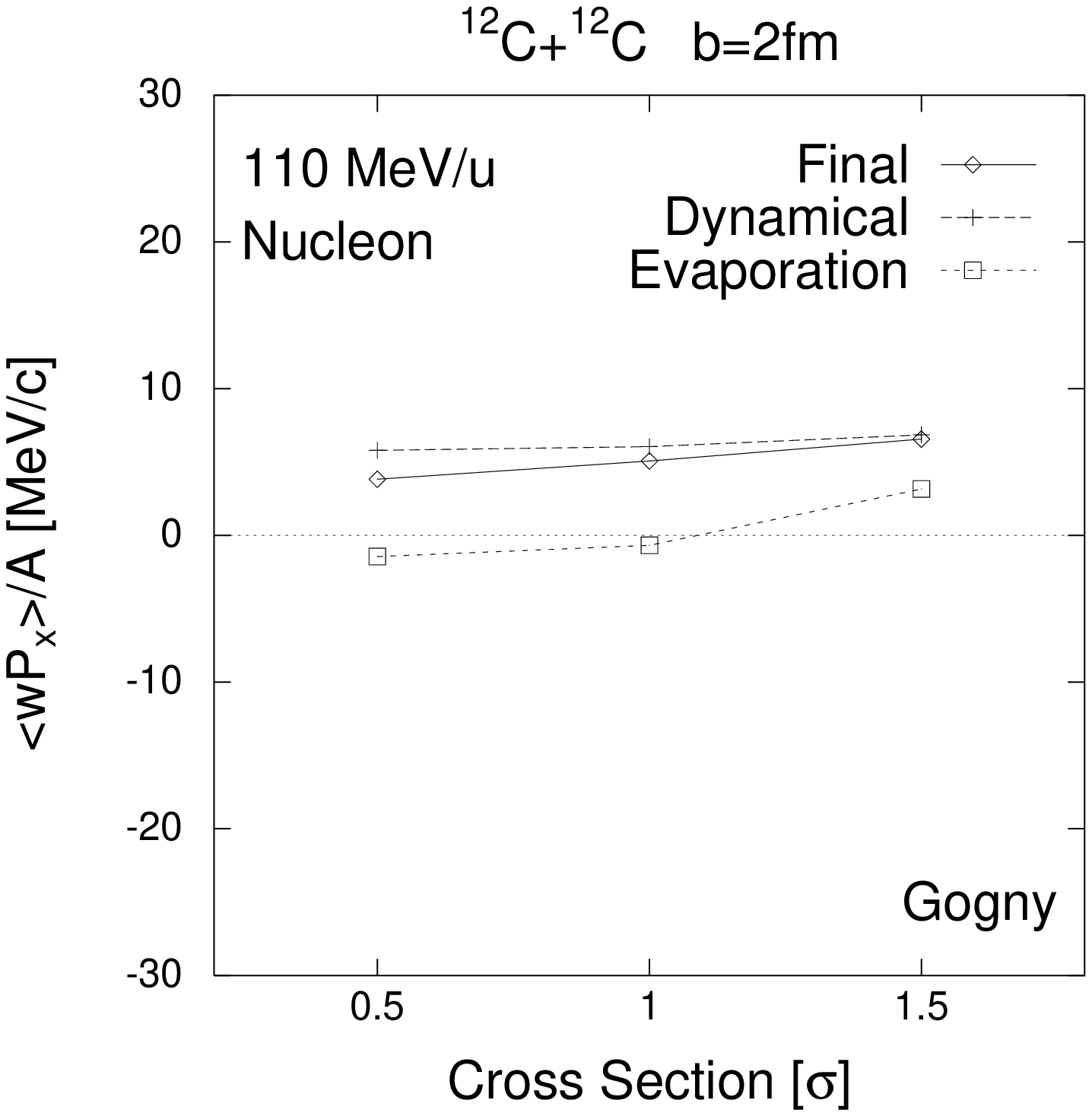}
            \epsfysize=9cm \epsffile{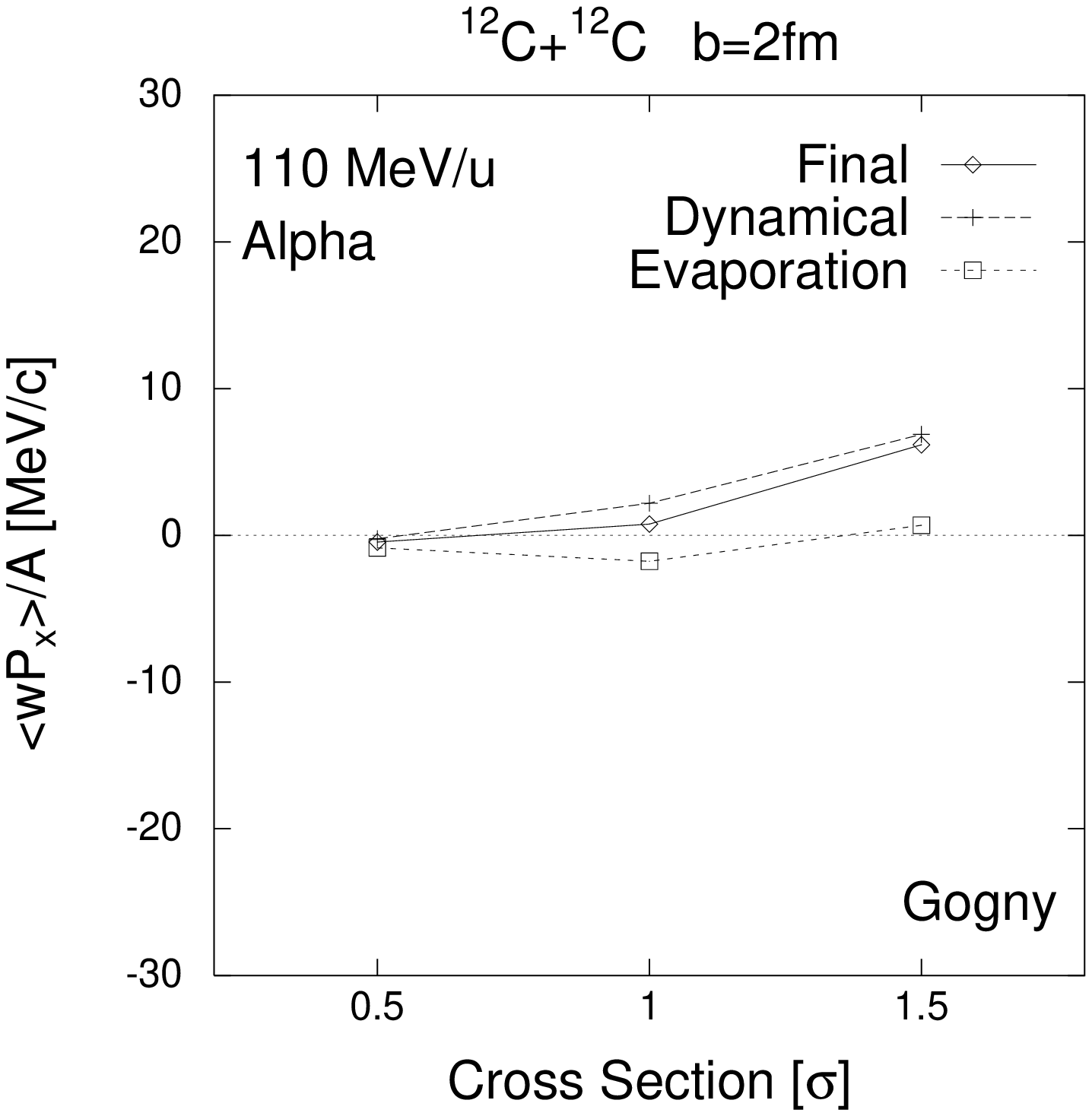}}
\caption{\label{FigFlowSig2}
Dependence of the flows on the stochastic collision cross sections for
the reaction $\carbon+\carbon$ at 110 MeV/nucleon.  }
\end{figure}

Figs.\ \ref{FigFlowSig} and \ref{FigFlowSig2} show the
$\sigma$-dependence of the flows of nucleons and $\alpha$ particles at
the incident energy 45 MeV/nucleon and 110 MeV/nucleon, respectively,
when all the cross sections (see Appendix B) are multiplied by 0.5,
1.0 and 1.5.  The flows of the dynamically produced particles and the
flows of the particles produced by the statistical decay are shown
together with the final results.  The flow of the dynamically emitted
nucleons is almost independent of the cross sections.  The
multiplicity of dynamically emitted nucleons in the reaction at 45
MeV/nucleon is 2.00, 3.52 and 4.95 for $0.5\sigma$, $1.0\sigma$ and
$1.5\sigma$, respectively.  These results are consistent with the
interpretation in the previous section because what is important to
the flow of the dynamically emitted nucleons is that the emitted
nucleons are direct products of the stochastic collisions and the
number of such nucleons is not relevant to the value of the flow.

\begin{figure}
\centerline{\epsfysize=7cm \epsffile{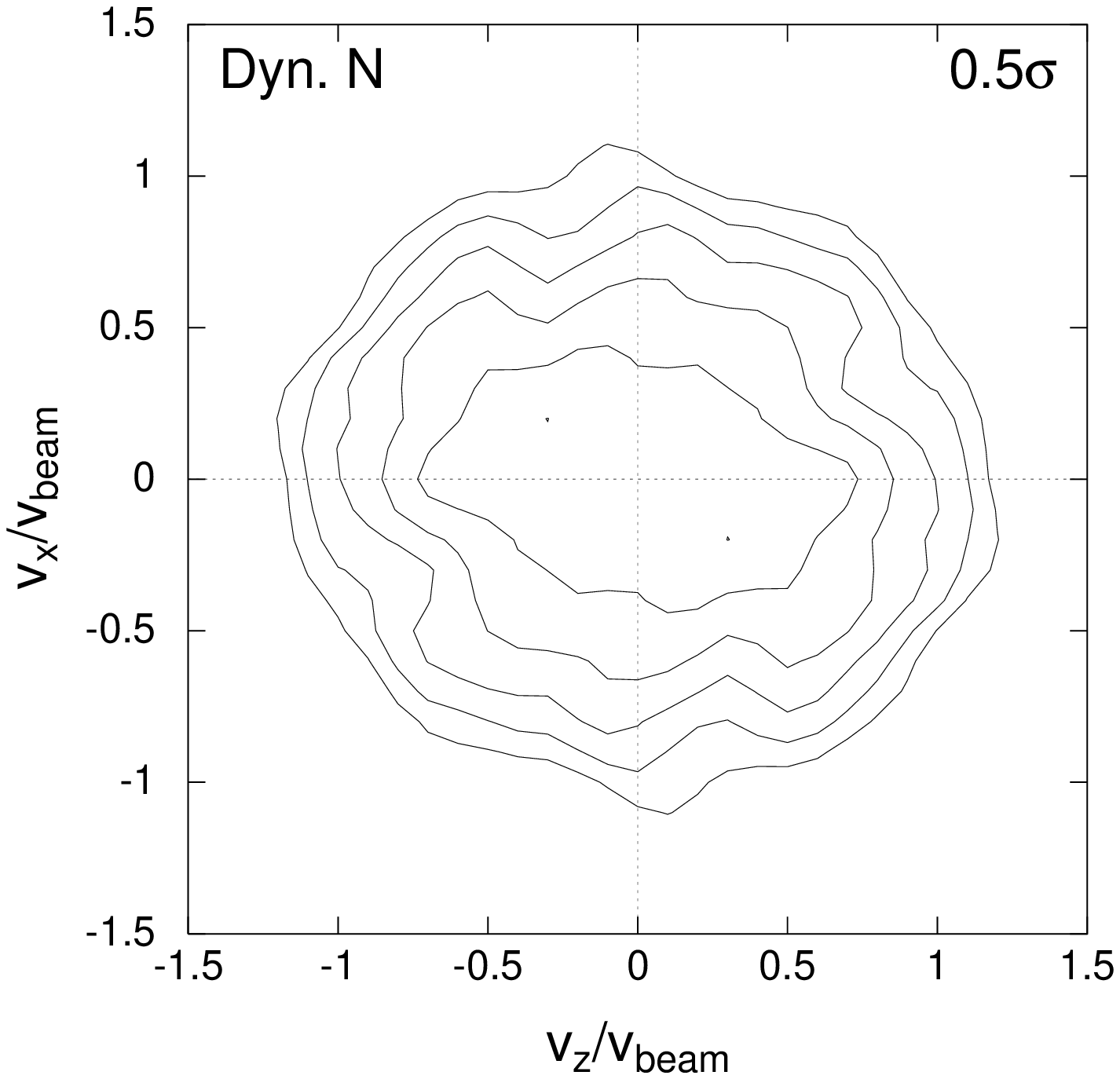}
            \epsfysize=7cm \epsffile{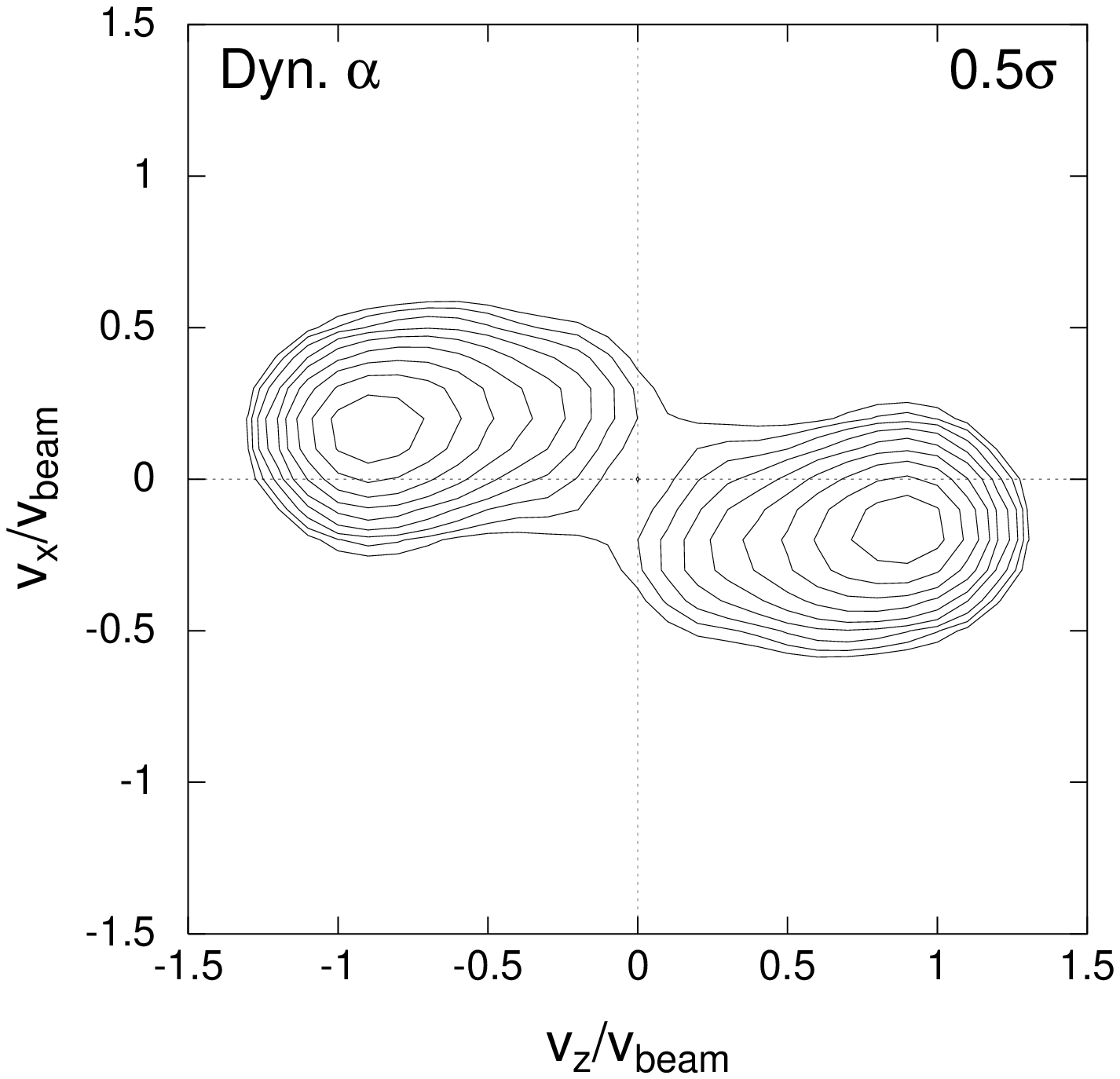}}\nobreak
\centerline{\epsfysize=7cm \epsffile{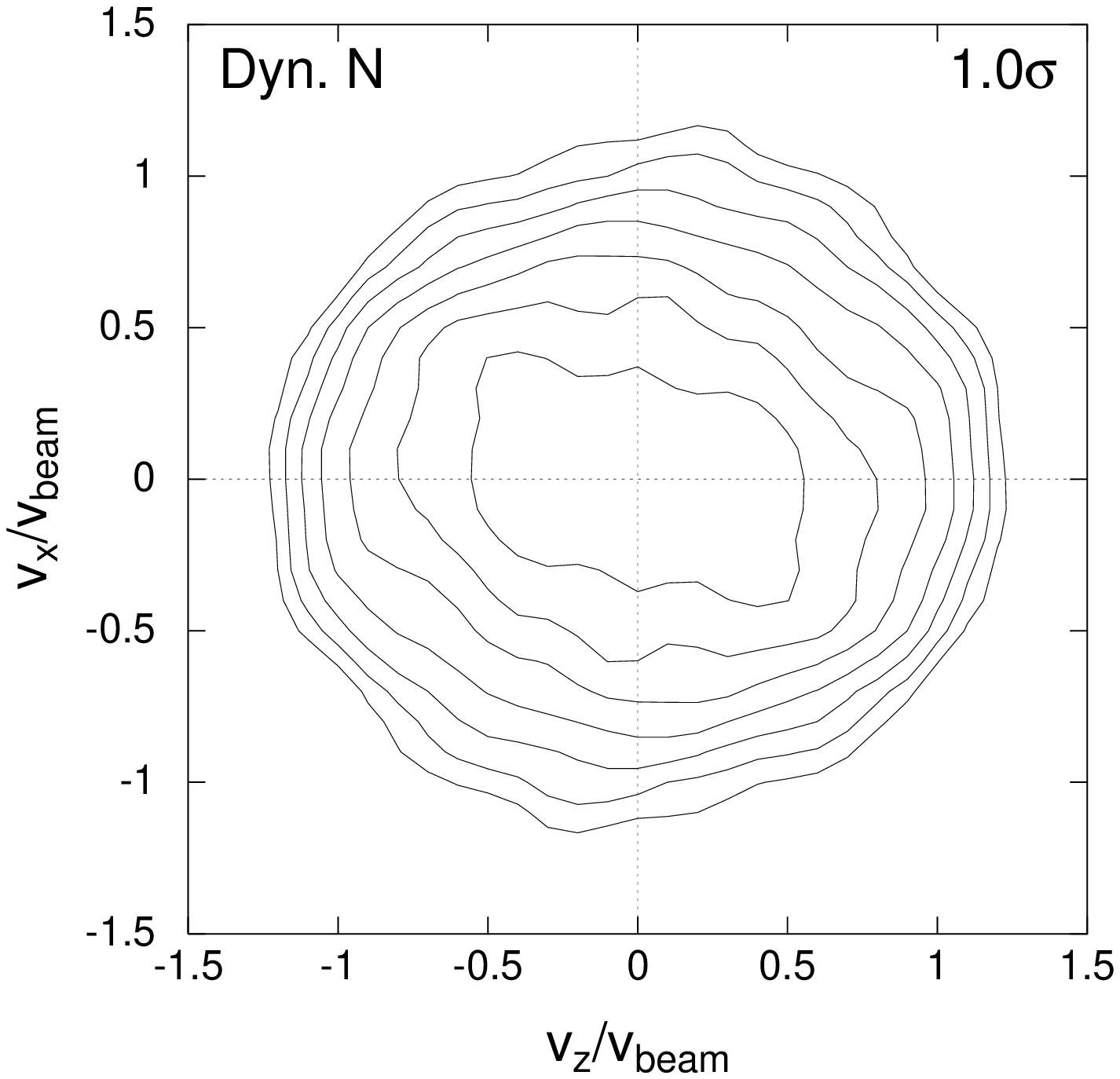}
            \epsfysize=7cm \epsffile{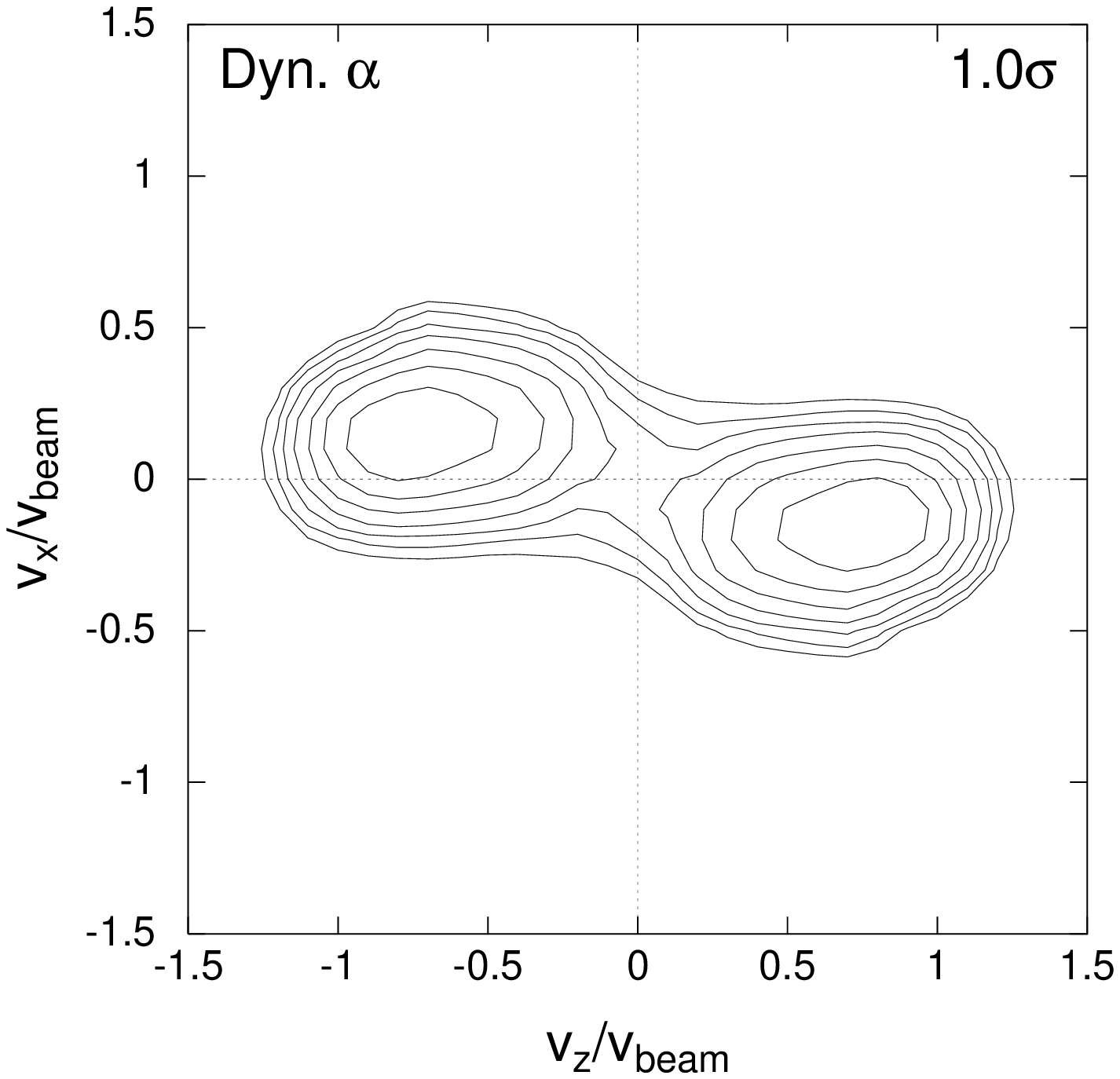}}\nobreak
\centerline{\epsfysize=7cm \epsffile{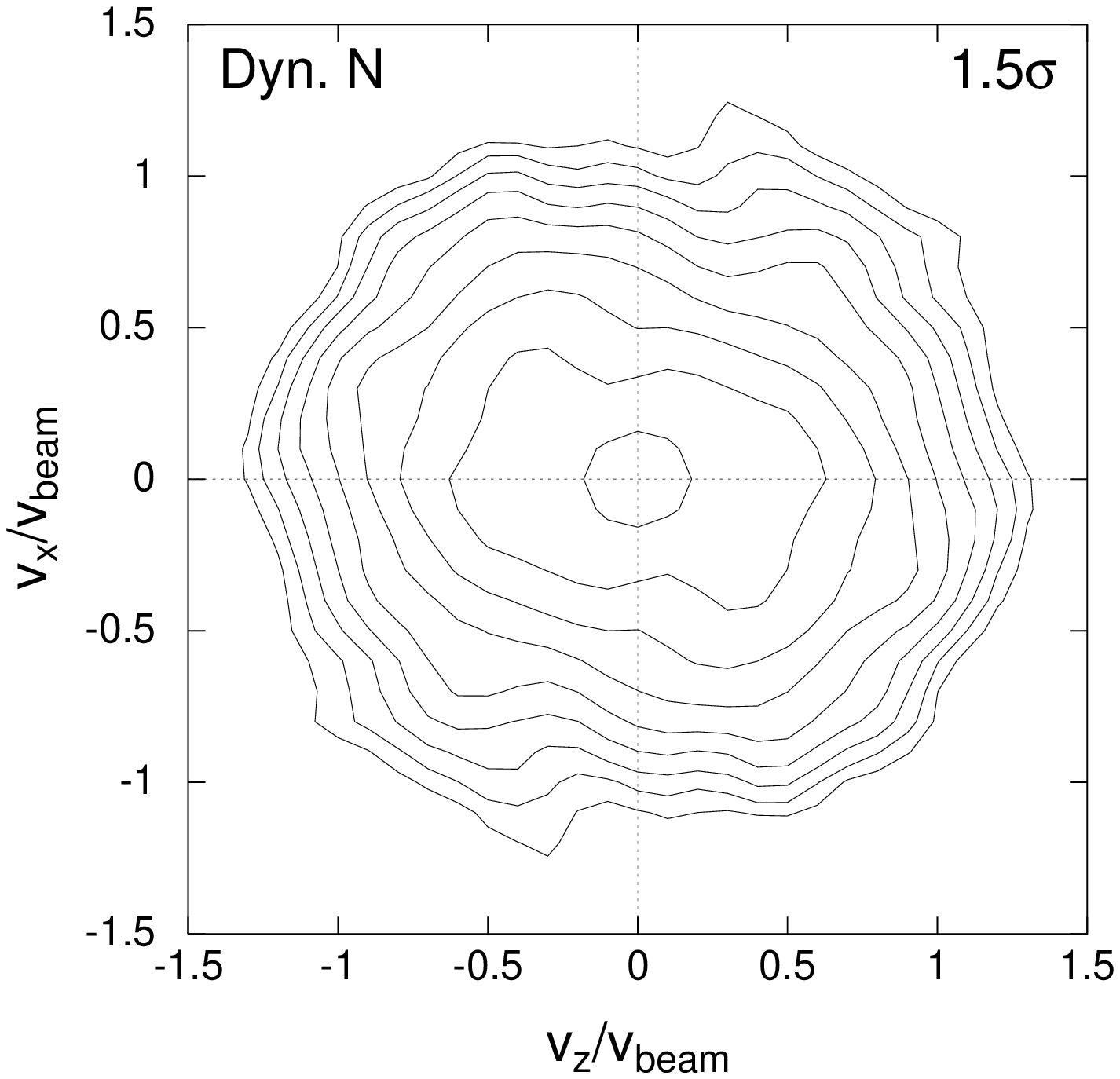}
            \epsfysize=7cm \epsffile{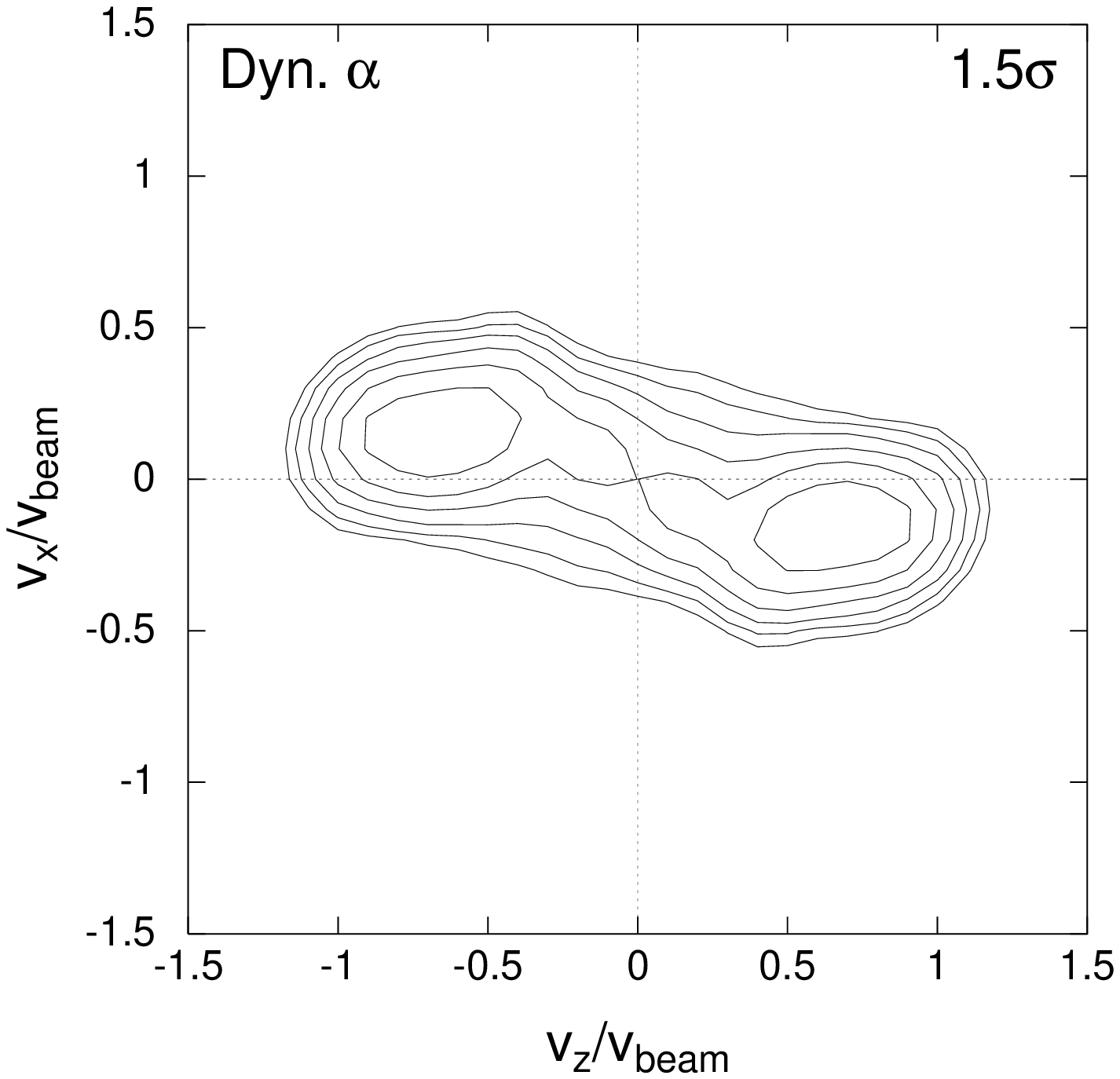}}
\caption{\label{FigMomDistDyn}
Momentum distributions of nucleons and $\alpha$ particles at $t=150$
fm/$c$ before the statistical decay in the reaction ${}^{12}{\rm
C}+{}^{12}{\rm C}$ at 45 MeV/nucleon with $b=2$ fm.  Results for the
three sets $0.5\sigma$, $1.0\sigma$ and $1.5\sigma$ of the stochastic
collision cross sections are shown.  }
\end{figure}

\begin{figure}
\centerline{\epsfysize=7cm \epsffile{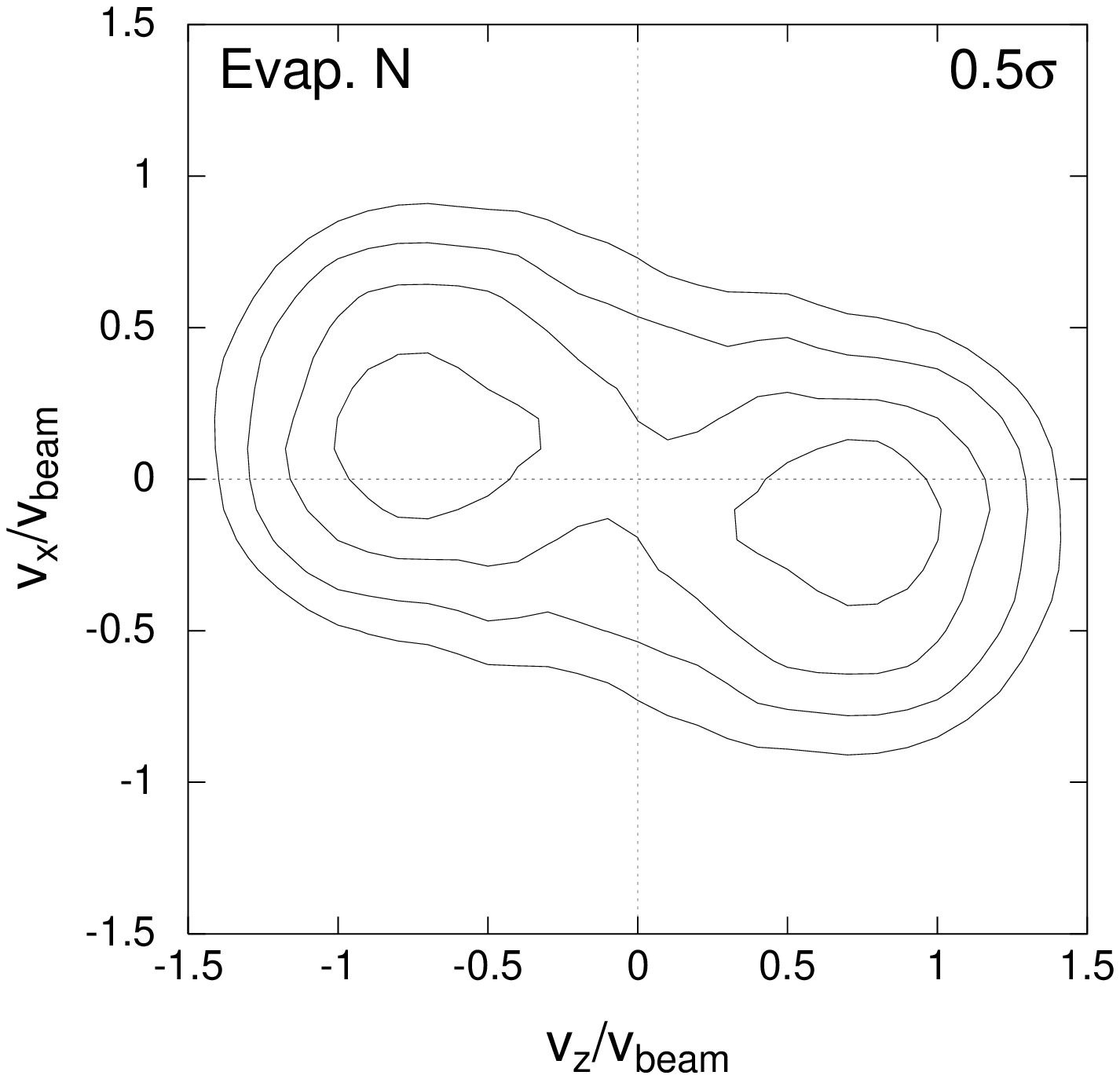}
            \epsfysize=7cm \epsffile{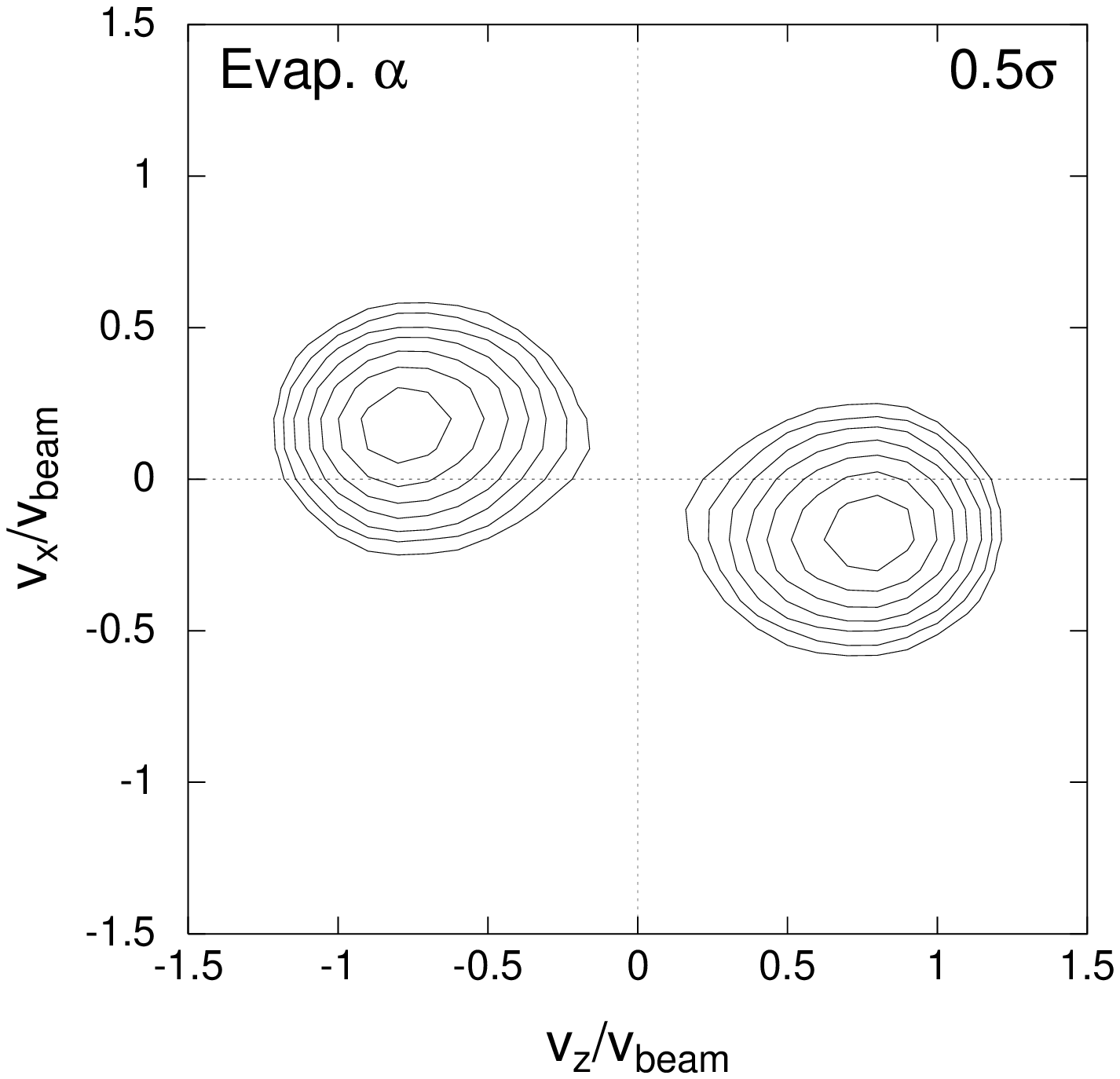}}\nobreak
\centerline{\epsfysize=7cm \epsffile{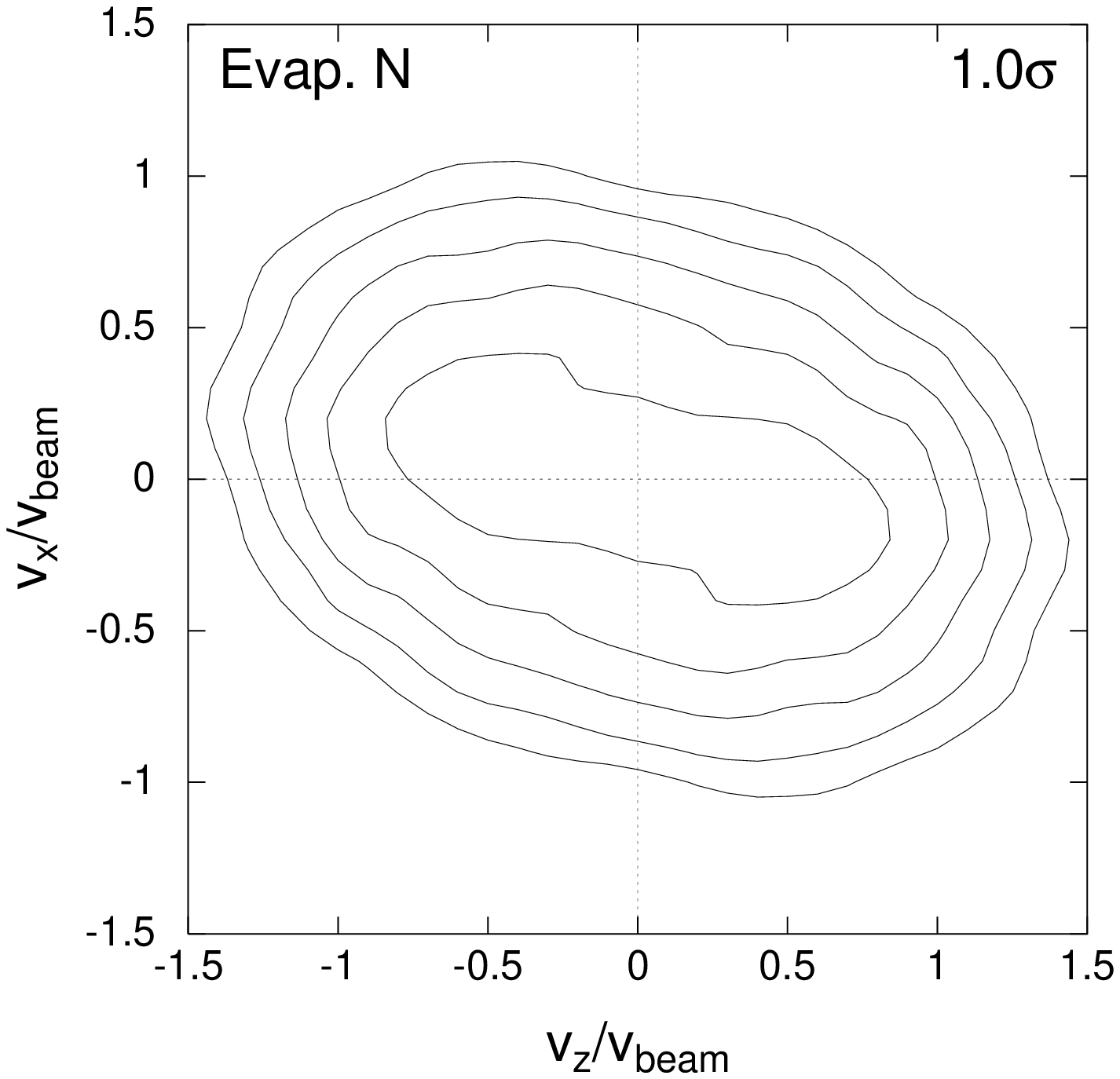}
            \epsfysize=7cm \epsffile{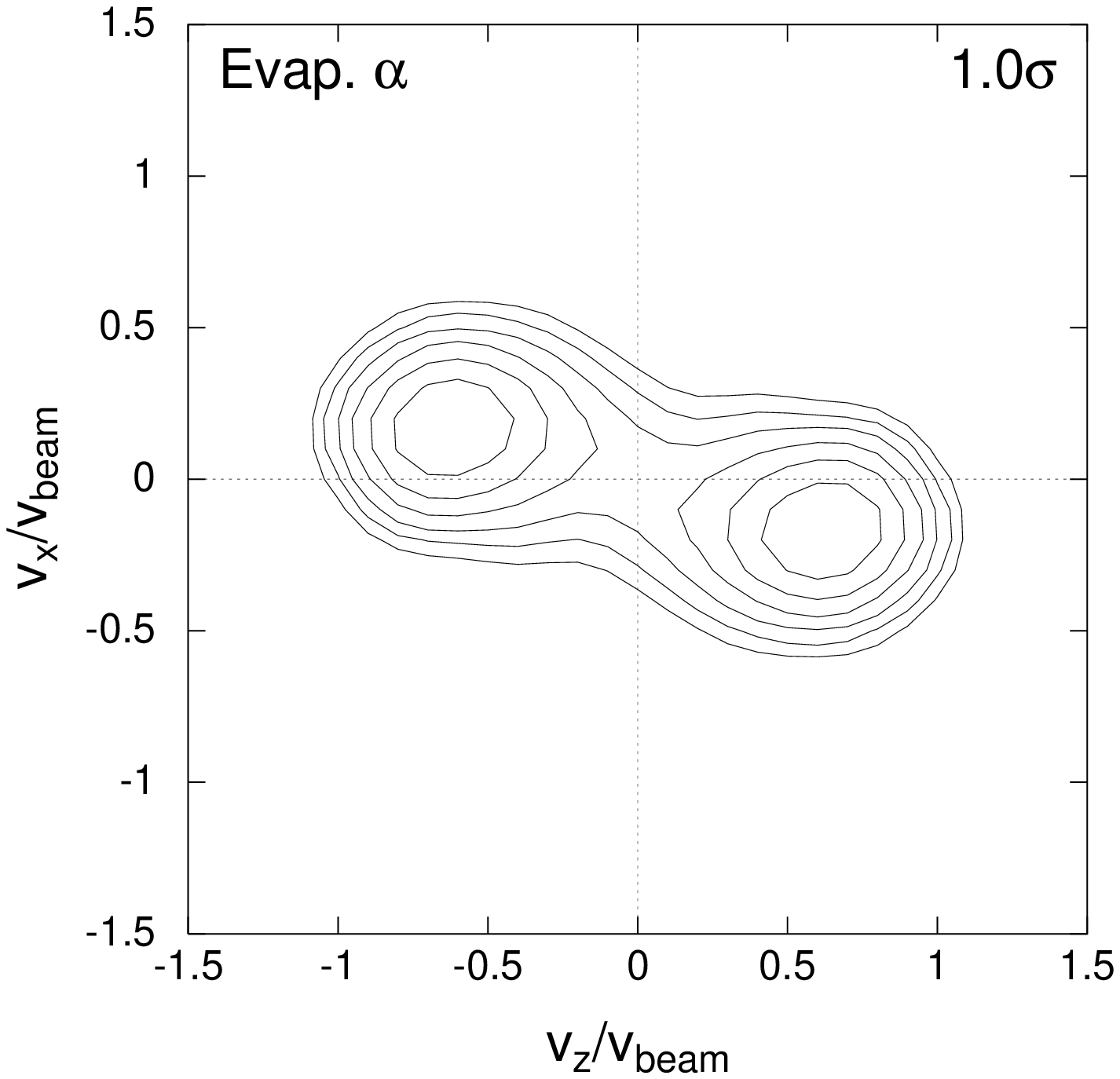}}\nobreak
\centerline{\epsfysize=7cm \epsffile{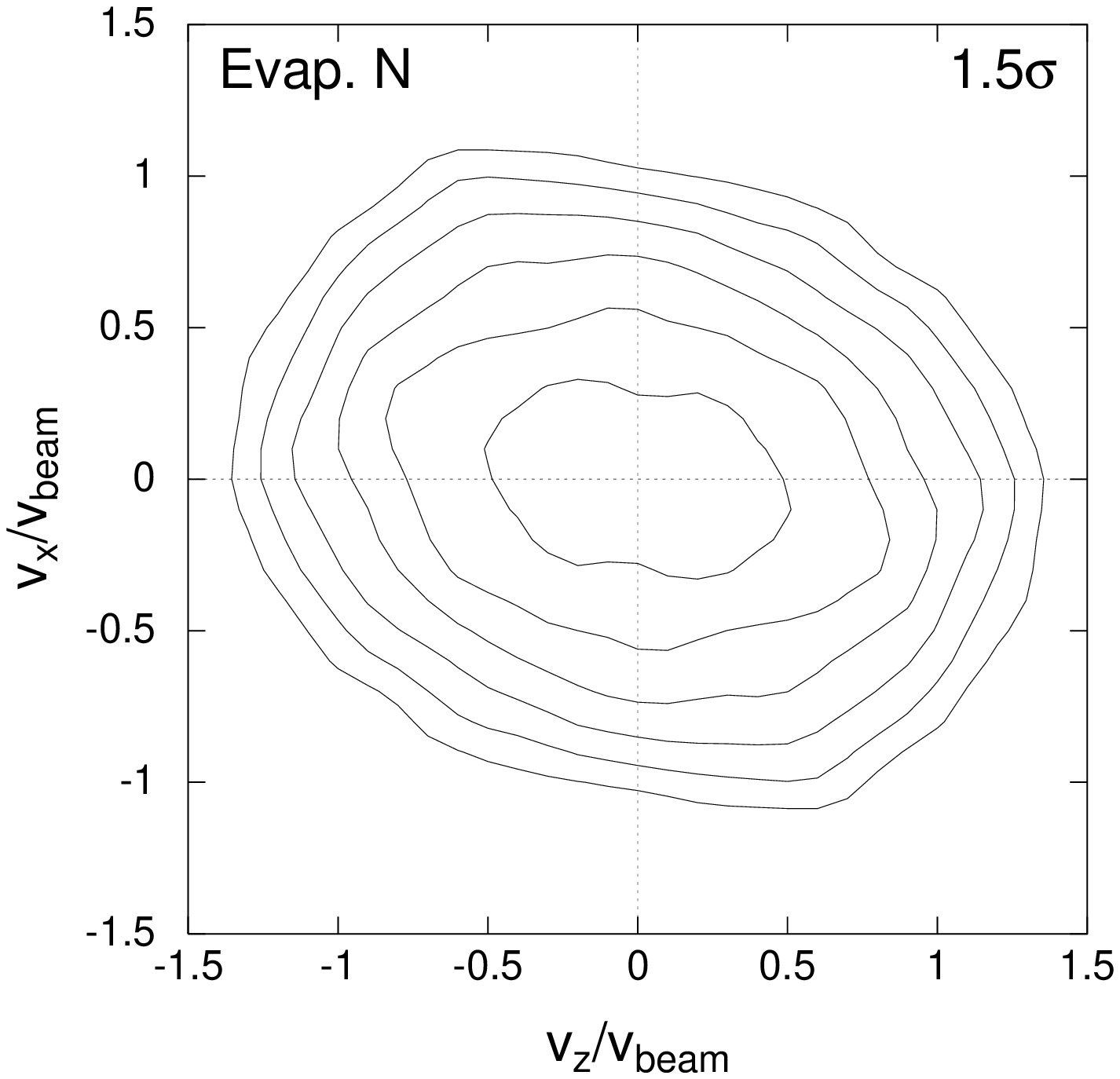}
            \epsfysize=7cm \epsffile{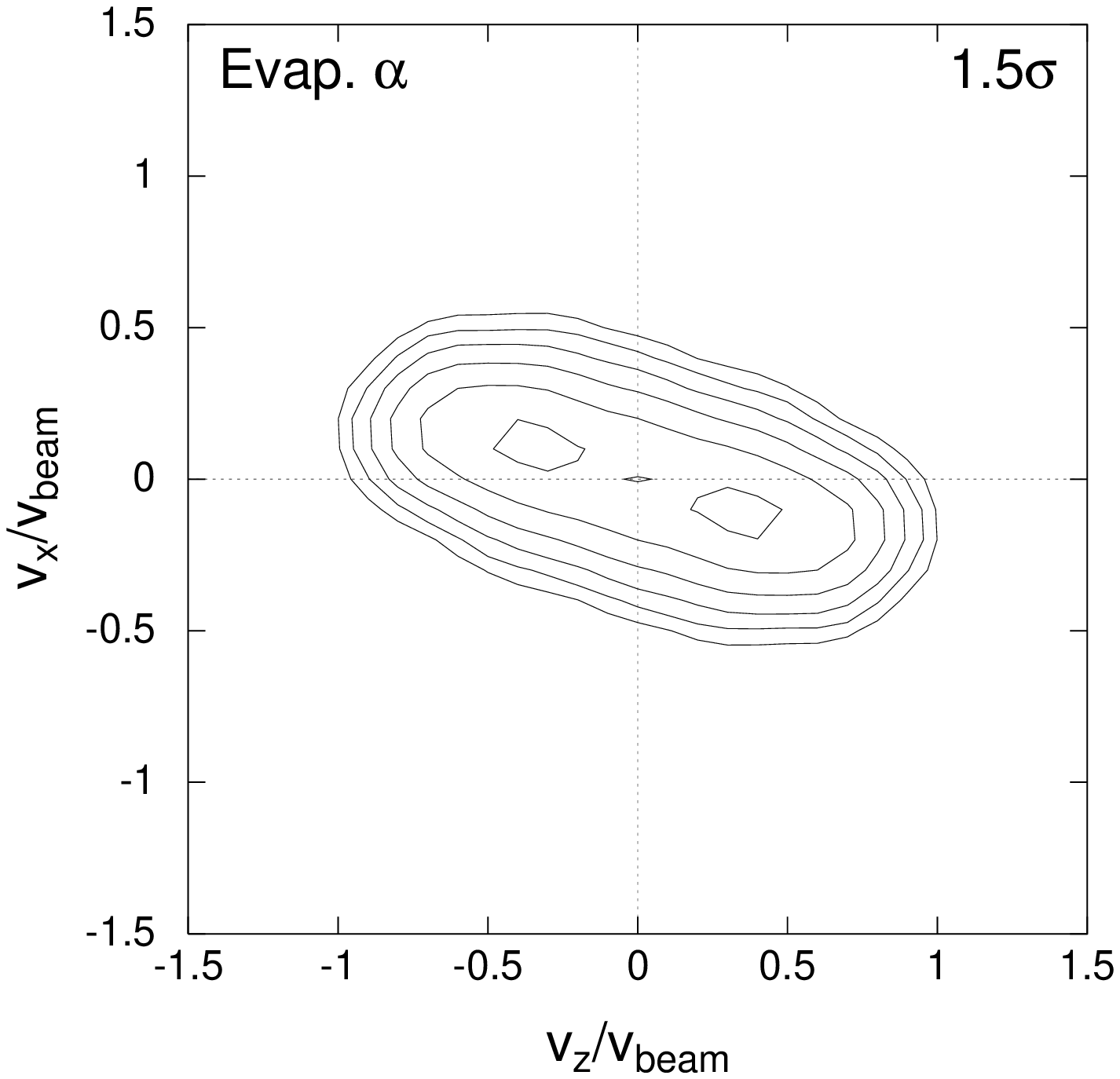}}
\caption{\label{FigMomDistEvap}
The same as Fig.\ \protect\ref{FigMomDistDyn} but calculated from only
the particles produced by the statistical decay.  }
\end{figure}

On the other hand, the $\sigma$-dependence of the flow of the nuclear
matter is large, especially in the reaction at 45 MeV/nucleon, and the
absolute value of the negative flow is smaller for larger cross
sections.  This might sound inconsistent to the naive expectation from
the previous interpretation that the flow of the matter is essentially
affected by the mean field.  This $\sigma$-dependence, however, turns
out to be very reasonable when one pays attention to other features of
the momentum distribution as well as the flow.  Figs.\
\ref{FigMomDistDyn} and \ref{FigMomDistEvap} show the in-plane
momentum distribution of nucleons and $\alpha$ particles which are
calculated at $t=150$ fm/$c$ (Dynamical) and from those produced by
the statistical decay process (Evaporation), respectively.  Results
for three cross sections $0.5\sigma$, $1.0\sigma$ and $1.5\sigma$ are
shown for the reaction at 45 MeV/nucleon.  The momentum distribution
of dynamically emitted nucleons is almost spherical while the other
three momentum distributions (of evaporated nucleons, dynamical
$\alpha$ particles, and evaporated $\alpha$ particles) have the common
$\sigma$-dependence though the momentum distribution of the evaporated
nucleons has wider spreading compared to that of $\alpha$ particles.
As the cross sections are increased, the dissipated component of the
momentum distribution around the center-of-mass momentum increases
while the flow angle does not change.  What has changed the flow value
of the matter is nothing but this $\sigma$-dependence of the
dissipated component of the momentum distribution.

\begin{figure}
\centerline{\epsfysize=9cm \epsffile{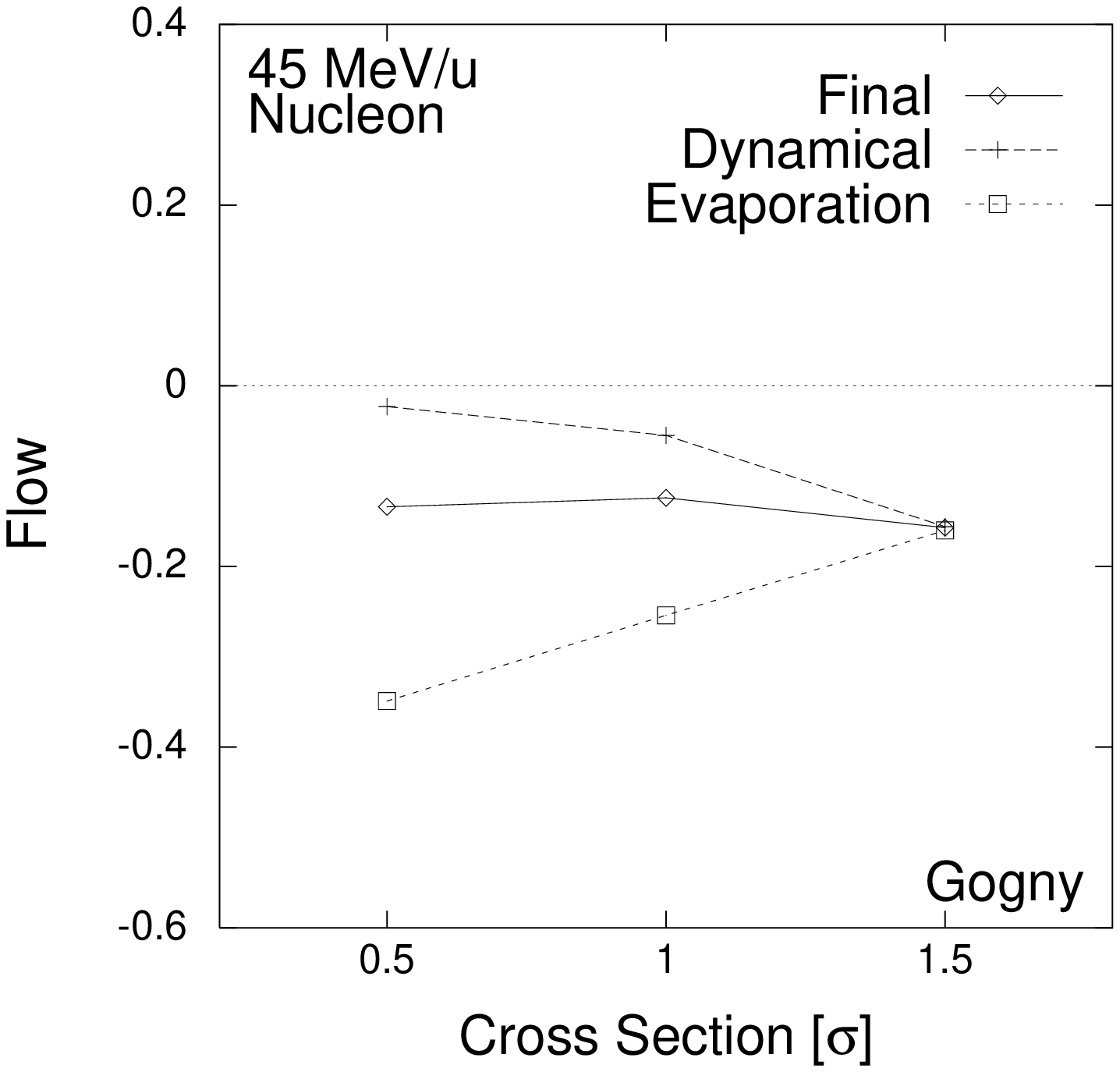}}
\caption{\label{FigFlow2Sig}
Dependence of the flow of nucleons on the stochastic collision cross
sections for the reaction $\carbon+\carbon$ at 45 MeV/nucleon.
The flow in this figure is defined as the slope of $\langle
P_x\rangle$-$P_z$ curve at $P_z=0$.  }
\end{figure}

It is to be noted here that the $\sigma$-dependence of the flow is not
due to the specific definition of the flow (\ref{EqDefFlow}).  For
example, if we adopt the definition of the flow as the slope of
$\langle P_x\rangle$-$P_z$ curve at $P_z=0$, we get the
$\sigma$-dependence of the nucleon flow shown in Fig.\
\ref{FigFlow2Sig} which has the same qualitative feature as the left
part of Fig.\ \ref{FigFlowSig}.

\section{Summary}
In this paper the collective flow of fragments was calculated as well
as the flow of nucleons with the AMD for the reaction ${}^{12}{\rm
C}+{}^{12}{\rm C}$ in the energy range 45 MeV/nucleon $\le E\le$ 150
MeV/nucleon.  The balance energy observed in the experiment was found
to be consistent with the Gogny force which yields a soft equation of
state with $K=228$ MeV.  We obtained larger absolute value of the
fragment flow than the nucleon flow, which is also consistent with the
experimentally observed feature.

The energy dependence and the $\sigma$-dependence of the calculated
flows of nucleons and fragments were found to be understandable, at
least in the energy region where the flow is negative, by introducing
the idea of two components of the collective flow.  The first
component is the flow of the dynamically emitted nucleons which are
produced by stochastic collisions in the early stage of the reaction.
Since the stochastic collisions erase the effect of the mean field,
the flow of the dynamically emitted nucleons is small.  The second
component is the flow of the nuclear matter which contributes to the
flows of fragments produced during the dynamical stage of the
reaction. The flows of the evaporated nucleons and fragments inherit
the flow of the parent fragments and hence the flows of dynamical
fragments and the evaporated nucleons and fragments have almost common
magnitude and common energy dependence and $\sigma$-dependence.
Although the second component is affected largely by the mean field,
its $\sigma$-dependence is also large.  However, this
$\sigma$-dependence turned out to come not from the change of the flow
angle but from change of the yield of the dissipated component in the
momentum distribution of the fragments.  Detailed comparison of the
fragment momentum distribution with the data will be able to settle
the in-medium cross sections.  On the other hand, nucleon momentum
distribution is mainly governed by the dynamically emitted nucleons in
this energy region, and therefore the shape of the momentum
distribution is not so sensitive to the in-medium cross sections.

We have mainly concentrated on the mechanism of the flow creation and
the $\sigma$-dependence in this paper, but the dependence on the
effective interaction was not discussed in detail.  What we have found
is that the Gogny force is sufficient for the reproduction of the
balance energy while the Volkov force fails.  To calculate the flow
with other effective interactions which reproduce the saturation
property of the nuclear matter and have various incompressibilities is
an important future problem for the determination of the equation of
state of the nuclear matter.

\acknowledgments
The computational calculation for this work has been financially
supported by Research Center for Nuclear Physics, Osaka University, as
an RCNP Computational Nuclear Physics Project (Project No.\
92--B--04).

\appendix
\section{Density dependent interaction}
Here we consider an effective interaction which contains density
dependent zero-range part,
\begin{equation}
V_{\rho}={t_\rho\over6}\sum_{i<j}X_{ij}\rho^\sigma(\vec r_i)
                                 \delta(\vec r_i-\vec r_j).
\end{equation}
For example, the density dependent part of the Gogny force has this
form with
\begin{equation}
t_\rho=6t_3,\quad X=1+P_\sigma,\quad \sigma=1/3.
\end{equation}
By using the density $\rho(\vec r)$ which is calculated from the AMD
wave function, the expectation value of $V_\rho$ can be written as
\begin{equation}
{\cal V}_\rho=\langle V_\rho\rangle
     = {t_\rho\over16}\int d\vec r\,\mu(\vec r)\rho^\sigma(\vec r),
\label{EqVrho}
\end{equation}
where
\begin{eqnarray}
&&
\mu(\vec r)=\sum_\alpha\tilde\rho_\alpha(\vec r)\rho_\alpha(\vec r),
\qquad
\rho(\vec r)=\sum_\alpha\rho_\alpha(\vec r),
\\&&
\rho_\alpha(\vec r)=\Bigl({2\nu\over\pi}\Bigr)^{3/2}
         \sum_{i,j\in\alpha}
         e^{-2\bigl(\sqrt\nu\vec r-(\vec Z_i^*+\vec Z_j)/2\bigr)^2}
         B_{ij}B^{-1}_{ji},
\\&&
\tilde\rho_\alpha(\vec r)=\sum_\beta
{4\over3}\langle\alpha\beta|X|\alpha\beta-\beta\alpha\rangle
\rho_\beta(\vec r).
\end{eqnarray}
Since the analytical integration of (\ref{EqVrho}) is impossible for
non-integer $\sigma$ and even in the case of $\sigma=1$ the evaluation
of the analytical six fold summation is not feasible, we evaluate
${\cal V}_\rho$ by the Monte Carlo method generating $N_{\rm TP}$ test
particles randomly with an appropriate weight function $\mu_{\rm
w}(\vec r)$,
\begin{equation}
{\cal V}_\rho\approx {t_\rho\over16}{F\over N_{\rm TP}}
     \sum_{k=1}^{N_{\rm TP}}{\mu(\vec r_k)\over\mu_{\rm w}(\vec r_k)}
     \rho^\sigma(\vec r_k),
\quad
F=\int d\vec r\,\mu_{\rm w}(\vec r),
\end{equation}
where $\vec r_k$ is the position of the $k$th test particle.  The
arbitrary function $\mu_{\rm w}(\vec r)$ should be chosen so as to
cover the important region in the integral (\ref{EqVrho}) efficiently
and at the same time to make the generation of test particles
numerically easy.  By using the approximate density $\rho_{\rm w}(\vec
r)$ calculated from the physical positions $\vec R_i={\rm Re}\vec W_i$
as
\begin{equation}
\rho_{{\rm w}}(\vec r)=\Bigl({2\nu\over\pi}\Bigr)^{3/2}
                    \sum_{i}e^{-2(\sqrt\nu\vec r-\vec R_i)^2},
\end{equation}
we take $\mu_{\rm w}(\vec r)$ as
\begin{eqnarray}&&
\mu_{\rm w}(\vec r)=\rho_{{\rm w}}(\vec r)^2
=\Bigl({2\nu\over\pi}\Bigr)^{3}
  \sum_{ij}
  e^{-(\vec R_i-\vec R_j)^2}
  e^{-4\bigl(\sqrt\nu\vec r-(\vec R_i+\vec R_j)/2\bigr)^2},
\\&&
F =\Bigl({\nu\over\pi}\Bigr)^{3/2}
  \sum_{ij}e^{-(\vec R_i-\vec R_j)^2}.
\end{eqnarray}

Derivatives of ${\cal V}_\rho$ with respect to $\vec Z_h$ are
necessary to solve the equation of motion of AMD.
They are also evaluated by the Monte Carlo method as
\begin{eqnarray}
{\partial{\cal V}_\rho\over\partial\vec Z_h^*}
&=&{t_\rho\over16}\int d\vec r\,
  \Bigl[2\tilde\rho_\alpha(\vec r)
  +{\sigma\mu(\vec r)\over\rho(\vec r)}\Bigr]
  \rho^\sigma(\vec r)
  {\partial\rho_\alpha(\vec r)\over\partial\vec Z_h^*}
\\
&\approx&{t_\rho\over16}{F\over N_{\rm TP}}\sum_{k=1}^{N_{\rm TP}}
        \Bigl[2\tilde\rho_\alpha(\vec r_k)
              +{\sigma\mu(\vec r_k)\over\rho(\vec r_k)}\Bigr]
        {\rho^\sigma(\vec r_k)\over\mu_{\rm w}(\vec r_k)}
        {\partial\rho_\alpha(\vec r_k)\over\partial\vec Z_h^*},
\end{eqnarray}
where $\alpha$ is the spin-isospin label of the nucleon $h$.

\section{Detail of the stochastic collision process}
In the higher energy region, we use the proton-neutron and
proton-proton collision cross sections parametrized as
\begin{eqnarray}&&
\sigma_{pn}=
\max\Bigl\{13335\,E{\rm[MeV]}^{-1.125}, 40\Bigr\}\,{\rm mb},
\label{EqSigpnhigh}
\\&&
\sigma_{pp}=
\max\Bigl\{4445\,E{\rm[MeV]}^{-1.125}, 25\Bigr\}\,{\rm mb},
\label{EqSigpphigh}
\end{eqnarray}
where $E$ is the energy in the laboratory system of the stochastic
collisions.  These are based on the data of free cross sections.  The
neutron-neutron collision cross section is assumed to be identical to
the proton-proton collision cross section.  The nucleon-alpha
collision total cross section $\sigma_{N\alpha,{\rm tot}}$ is chosen
to be equal to the total cross section of the process where the
nucleon were scattered by two-nucleon collisions with four nucleons in
the $\alpha$ cluster with the cross sections $\sigma_{pn}$ and
$\sigma_{pp}$.  The adopted nucleon-alpha inelastic collision cross
section is based on the experimental data parametrized \cite{ONOc} as
\begin{equation}
\sigma_{N\alpha,{\rm inel}}=\max\Bigl\{
120-162e^{-(E-20\,{\rm MeV})/(10\,{\rm MeV})}, 0\Bigr\}\,{\rm mb} .
\end{equation}
The angular distribution of proton-neutron scatterings are taken as
\begin{equation}
{d\sigma_{pn}\over d\Omega}\propto
10^{-\alpha\bigl(\pi/2-|\theta-\pi/2|\bigr)},
\quad
\alpha={2\over\pi}\max\Bigl\{0.333\ln E{\rm [MeV]}-1,\, 0\Bigr\},
\end{equation}
while the proton-proton and neutron-neutron scatterings are assumed to
be isotropic.  The nucleon-alpha inelastic collision is treated as a
two-nucleon collision between the nucleon and a nucleon in the
$\alpha$ cluster and the same angular distribution as the two-nucleon
collision is applied.  The angular distribution of the nucleon-alpha
elastic scattering is parametrized \cite{ONOc} as
\begin{equation}
{d\sigma_{N\alpha,\rm el}\over d\Omega}\propto
\exp\biggl[-\Bigl({180^\circ-\theta\over70^\circ}\Bigr)^2\biggl]
+10\exp\biggl[-\Bigl({\theta-20^\circ\over40^\circ}\Bigr)^2\biggl].
\end{equation}

Since the above defined cross sections are too large if they are
applied to low energy region, we use the cross sections
\begin{eqnarray}&&
\sigma_{pn}=\sigma_{pp}
={100\,{\rm mb}\over
1+E/(200\,{\rm MeV})+2\min\{(\rho/\rho_0)^{1/2},1\}},
\label{EqSigNNlow}
\\&&
\sigma_{N\alpha,{\rm tot}}
={571\,{\rm mb}\over
1+E/(200\,{\rm MeV})+2\min\{(\rho/\rho_0)^{1/2},1\}}
\label{EqSigNalow}
\end{eqnarray}
which are similar to those we used in Ref.\ \cite{ONOc} for the
reaction at 28.7 MeV/nucleon.  The cross sections (\ref{EqSigNNlow})
and (\ref{EqSigNalow}) are used if they are smaller than the cross
sections $\sigma_{pn}$, $\sigma_{pp}$ and $\sigma_{N\alpha,{\rm tot}}$
defined for the higher energy region by Eqs.\ (\ref{EqSigpnhigh}) and
(\ref{EqSigpphigh}).  On the other hand, the same expressions of
angular distributions and $\sigma_{N\alpha,{\rm inel}}$ are used for
the whole energy region.

\section{Subtraction of spurious zero-point oscillation}
Like in Refs.\ \cite{ONOa,ONOb,ONOc} the spurious kinetic energies of
zero-point oscillations of center-of-mass motion of fragments are
subtracted by modifying the Hamiltonian in the equation of motion of
AMD as
\begin{equation}
{\cal H}=\langle H\rangle-{3\hbar^2\nu\over2M}A+T_0(A-N_{\rm F}),
\end{equation}
where the zero-point kinetic energy $T_0$ is treated as a free
parameter to adjust the gross feature of the binding energies of
nuclei.

We define the fragment number $N_{\rm F}$ as
\begin{equation}
  N_{\rm F}=\sum_{i=1}^A{g(k_i)\over n_im_i},
\end{equation}
where
\begin{equation}
  n_i=\sum_{j=1}^A \hat f_{ij}, \quad
  m_i=\sum_{j=1}^A {1\over n_j}f_{ij}, \quad
  k_i=\sum_{j=1}^A \bar f_{ij}
\end{equation}
and
\begin{eqnarray}&&
\hat f_{ij}=F(d_{ij},\hat\xi,\hat a),\quad
f_{ij}=F(d_{ij},\xi,a),\quad
\bar f_{ij}=F(d_{ij},\bar\xi,\bar a),
\\&&
d_{ij}=|{\rm Re}(\vec Z_i-\vec Z_j)|,
\\&&
F(d,\xi,a)=\left\{\begin{array}{ll}
              1               & \mbox{if $d\le a$}\\
              e^{-\xi(d-a)^2} & \mbox{if $d>a$   }
           \end{array}\right. .
\end{eqnarray}
Compared to Ref.\ \cite{ONOa} where the Volkov force was used as the
effective interaction, we have newly introduced a function
\begin{equation}
g(k)=1+g_0e^{-(k-M)^2/2\sigma^2},
\end{equation}
which should be equal to 1 in principle.  We have introduced it in
order to remedy the situation that nuclei around $^{12}{\rm C}$ are
underbound while the binding energies of other nuclei such as $\alpha$
and $^{16}{\rm O}$ are almost reproduced in the calculation with the
Gogny force.  We have chosen the parameters as shown in Table
\ref{TabSubZero}.
Binding energies of nuclei are sensitive only to two parameters $T_0$
and $M$.
\begin{table}
\caption{\label{TabSubZero}
Parameters concerned with the subtraction of spurious zero-point
oscillations of fragments.  Parameters which are used with the Gogny
force in this paper are shown as well as those used with the Volkov
force No.\ 1 with $m=0.576$.  }
\begin{tabular}{lrrrrrrrrrrr}
Force & $\xi$ & $a$ & $\hat\xi$ & $\hat a$ & $\bar\xi$ & $\bar a$
& $g_0$ & $\sigma$ & $M$ & $T_0\quad$ \\
\tableline
Volkov &1.0&0.2&2.0&0.1& ---& ---&---&---&---&7.7 MeV\\
Gogny  &2.0&0.6&2.0&0.2&1.0&0.5&1.0&2.0&12.0&9.2 MeV\\
\end{tabular}
\end{table}

\end{document}